\documentclass[twocolumn]{aastex63}
\usepackage{url}
\usepackage{threeparttable}

\usepackage{natbib}
\usepackage{appendix}
\usepackage{amsmath}
\received{August 11, 2021}
\revised{October 26, 2021}
\accepted{October 29, 2021}
\submitjournal{ApJ in Press}

\shorttitle{Extremely Metal-Poor Low-Mass Galaxies and HN/PISN Signatures}
\shortauthors{Isobe et al.}
\graphicspath{{./}{figures/}}
\def\tcra{\textcolor{black}}
\def\tcrb{\textcolor{black}}
\def\tcrc{\textcolor{black}}
\def\tcrd{\textcolor{black}}
\def\tcre{\textcolor{black}}
\def\tcrf{\textcolor{black}}
\def\tcrg{\textcolor{black}}

\begin{document}

\title{EMPRESS. IV.\\
Extremely Metal-Poor Galaxies (EMPGs) Including Very Low-Mass\\
Primordial Systems with $M_{*}=10^{4}$--$10^{5}$ M$_{\odot}$ and 2--3\% (O/H)$_{\odot}$:\\ 
High (Fe/O) Suggestive of Metal Enrichment by Hypernovae/Pair-Instability Supernovae
}

\author[0000-0001-7730-8634]{Yuki Isobe}
\affiliation{Institute for Cosmic Ray Research, The University of Tokyo, 5-1-5 Kashiwanoha, Kashiwa, Chiba 277-8582, Japan}
\affiliation{Department of Physics, Graduate School of Science, The University of Tokyo, 7-3-1 Hongo, Bunkyo, Tokyo 113-0033, Japan}

\author[0000-0002-1049-6658]{Masami Ouchi}
\affiliation{National Astronomical Observatory of Japan, 2-21-1 Osawa, Mitaka, Tokyo 181-8588, Japan}
\affiliation{Institute for Cosmic Ray Research, The University of Tokyo, 5-1-5 Kashiwanoha, Kashiwa, Chiba 277-8582, Japan}
\affiliation{Kavli Institute for the Physics and Mathematics of the Universe (WPI), University of Tokyo, Kashiwa, Chiba 277-8583, Japan}

\author[0000-0002-7043-6112]{Akihiro Suzuki}
\affiliation{National Astronomical Observatory of Japan, 2-21-1 Osawa, Mitaka, Tokyo 181-8588, Japan}

\author[0000-0003-1169-1954]{Takashi J. Moriya}
\affiliation{National Astronomical Observatory of Japan, 2-21-1 Osawa, Mitaka, Tokyo 181-8588, Japan}
\affiliation{School of Physics and Astronomy, Faculty of Science, Monash University, Clayton, Victoria 3800, Australia}

\author[0000-0003-2965-5070]{Kimihiko Nakajima}
\affiliation{National Astronomical Observatory of Japan, 2-21-1 Osawa, Mitaka, Tokyo 181-8588, Japan}

\author[0000-0001-9553-0685]{Ken’ichi Nomoto}
\affiliation{Kavli Institute for the Physics and Mathematics of the Universe (WPI), University of Tokyo, Kashiwa, Chiba 277-8583, Japan}

\author{Michael Rauch}
\affiliation{Carnegie Observatories, 813 Santa Barbara Street, Pasadena, CA 91101, USA}

\author[0000-0002-6047-430X]{Yuichi Harikane} 
\affiliation{Institute for Cosmic Ray Research, The University of Tokyo, 5-1-5 Kashiwanoha, Kashiwa, Chiba 277-8582, Japan}
\affiliation{Department of Physics and Astronomy, University College London, Gower Street, London WC1E 6BT, UK}

\author[0000-0001-5780-1886]{Takashi Kojima}
\affiliation{Institute for Cosmic Ray Research, The University of Tokyo, 5-1-5 Kashiwanoha, Kashiwa, Chiba 277-8582, Japan}
\affiliation{Department of Physics, Graduate School of Science, The University of Tokyo, 7-3-1 Hongo, Bunkyo, Tokyo 113-0033, Japan}

\author[0000-0001-9011-7605]{Yoshiaki Ono}
\affiliation{Institute for Cosmic Ray Research, The University of Tokyo, 5-1-5 Kashiwanoha, Kashiwa, Chiba 277-8582, Japan}

\author[0000-0001-7201-5066]{Seiji Fujimoto} 
\affiliation{Cosmic DAWN Center}
\affiliation{Niels Bohr Institute, University of Copenhagen, Lyngbyvej2, DK-2100, Copenhagen, Denmark}
\affiliation{Research Institute for Science and Engineering, Waseda University, 3-4-1 Okubo, Shinjuku, Tokyo 169-8555, Japan}
\affiliation{National Astronomical Observatory of Japan, 2-21-1 Osawa, Mitaka, Tokyo 181-8588, Japan}
\affiliation{Institute for Cosmic Ray Research, The University of Tokyo, 5-1-5 Kashiwanoha, Kashiwa, Chiba 277-8582, Japan}

\author[0000-0002-7779-8677]{Akio K. Inoue}
\affiliation{Waseda Research Institute for Science and Engineering, Faculty of Science and Engineering, Waseda University, 3-4-1, Okubo, Shinjuku, Tokyo 169-8555, Japan}
\affiliation{Department of Physics, School of Advanced Science and Engineering, Faculty of Science and Engineering, Waseda University, 3-4-1 Okubo, Shinjuku, Tokyo 169-8555, Japan}

\author[0000-0002-1418-3309]{Ji Hoon Kim}
\affiliation{Subaru Telescope, National Astronomical Observatory of Japan, National Institutes of Natural Sciences (NINS), 650 North Aohoku Place, Hilo, HI 96720, USA}
\affiliation{Metaspace, 36 Nonhyeon-ro, Gangnam-gu, Seoul 06312, Republic of Korea}

\author[0000-0002-3852-6329]{Yutaka Komiyama} 
\affiliation{National Astronomical Observatory of Japan, 2-21-1 Osawa, Mitaka, Tokyo 181-8588, Japan}
\affiliation{Department of Astronomical Science, SOKENDAI (The Graduate University for Advanced Studies), Osawa 2-21-1, Mitaka, Tokyo, 181-8588, Japan}

\author[0000-0002-3801-434X]{Haruka Kusakabe} 
\affiliation{Observatoire de Gen{\'e}ve, Universit{\'e} de Gen{\'e}ve, 51 Ch. des Maillettes, 1290 Versoix, Switzerland}

\author[0000-0003-1700-5740]{Chien-Hsiu Lee} 
\affiliation{NSF's National Optical-Infrared Astronomy Research Laboratory, 950 North Cherry Avenue, Tucson 85719, USA}

\author[0000-0003-0695-4414]{Michael Maseda}
\affiliation{Leiden Observatory, Leiden University, PO Box 9513, NL-2300 RA, Leiden, the Netherlands}

\author[0000-0003-2871-127X]{Jorryt Matthee}
\affiliation{Department of Physics, ETH Z{\"u}rich, Wolfgang-Pauli-Strasse 27, 8093 Z{\"u}rich, Switzerland}

\author{Leo Michel-Dansac}
\affiliation{Univ Lyon, Univ Lyon1, ENS de Lyon, CNRS, Centre de Recherche Astrophysique de Lyon UMR5574, 69230 Saint-Genis-Laval, France}

\author[0000-0002-7402-5441]{Tohru Nagao}
\affiliation{Research Center for Space and Cosmic Evolution, Ehime University, Matsuyama, Ehime 790-8577, Japan}

\author[0000-0003-2804-0648]{Themiya Nanayakkara}
\affiliation{Centre for Astrophysics \& Supercomputing, Swinburne University of Technology, PO Box 218, Hawthorn, VIC 3112, Australia}

\author{Moka Nishigaki}
\affiliation{Department of Astronomical Science, SOKENDAI (The Graduate University for Advanced Studies), Osawa 2-21-1, Mitaka, Tokyo, 181-8588, Japan}

\author[0000-0003-3228-7264]{Masato Onodera} 
\affiliation{Subaru Telescope, National Astronomical Observatory of Japan, National Institutes of Natural Sciences (NINS), 650 North Aohoku Place, Hilo, HI 96720, USA}
\affiliation{Department of Astronomical Science, SOKENDAI (The Graduate University for Advanced Studies), Osawa 2-21-1, Mitaka, Tokyo, 181-8588, Japan}

\author[0000-0001-6958-7856]{Yuma Sugahara} 
\affiliation{National Astronomical Observatory of Japan, 2-21-1 Osawa, Mitaka, Tokyo 181-8588, Japan}
\affiliation{Waseda Research Institute for Science and Engineering, Faculty of Science and Engineering, Waseda University, 3-4-1, Okubo, Shinjuku, Tokyo 169-8555, Japan}

\author{Yi Xu}
\affiliation{Institute for Cosmic Ray Research, The University of Tokyo, 5-1-5 Kashiwanoha, Kashiwa, Chiba 277-8582, Japan}
\affiliation{Department of Astronomy, Graduate School of Science, The University of Tokyo, 7-3-1 Hongo, Bunkyo, Tokyo 113-0033, Japan}

%% \collaboration{1}{(AAS Journals Data Scientists collaboration)}



\begin{abstract}
We present Keck/LRIS follow-up spectroscopy for 13 photometric candidates of extremely metal poor galaxies (EMPGs) selected by a machine-learning technique applied to the deep ($\sim 26$ AB mag) optical and wide-area ($\sim500$ deg$^2$) Subaru imaging data in the EMPRESS survey. 
Nine out of the 13 candidates are EMPGs with an oxygen abundance (O/H) less than $\sim10$\% solar value (O/H)$_\odot$, and four sources are contaminants of moderately metal-rich galaxies or no emission-line objects.
Notably, two out of the nine EMPGs have extremely-low stellar masses and oxygen abundances of $5\times10^{4}$--$7\times10^{5}$ M$_{\odot}$ and 2--3\% (O/H)$_\odot$, respectively. 
With a sample of five EMPGs with (Fe/O) measurements, two (three) of which are taken from this study (the literature), we confirm that \tcrg{two EMPGs with the lowest (O/H) ratios of $\sim2$\% (O/H)$_{\odot}$} show high (Fe/O) ratios of $\sim0.1$, close to the solar abundance ratio.
Comparing galaxy chemical enrichment models, we find that the \tcrg{two} EMPGs cannot be explained by a scenario of metal-poor gas accretion/episodic star-formation history due to their low (N/O) ratios.
We conclude that the \tcrg{two} EMPGs can be reproduced by an inclusion of bright hypernovae and/or hypothetical pair-instability supernovae (SNe) preferentially produced in a metal-poor environment. 
This conclusion implies that primordial galaxies at $z\sim10$ could have a high abundance of Fe that is not originated from Type Ia SNe with delays, and that Fe may not serve as a cosmic clock for primordial galaxies.
\end{abstract}

%\keywords{galaxies: formation --- galaxies: structure --- galaxies: star formation --- galaxies: dwarf}


\section{Introduction} \label{sec:intro}
Galaxies in the early formation phase are the keys to understanding galaxy formation and evolution. 
Young galaxies, especially galaxies with stellar ages below $\sim300$ Myr, are expected to be metal-poor, because low- and intermediate-mass stars cannot contribute to chemical enrichment \tcra{before} finishing lifetimes of $\sim300$ Myr as main-sequence stars. 
\citet{Wise2012a} predict that a first galaxy at $z\gtrsim7$ (corresponding to the stellar age of $\lesssim300$ Myr) has a halo mass of $10^{7}$--$10^{9}$ M$_{\odot}$, a stellar mass of $10^{4}$--$10^{6}$ M$_{\odot}$, and a metallicity of 0.1--1\% solar abundance.

In such an extremely low-metallicity environment, progenitor gas clouds of stars are also metal-poor. 
The metal-poor gas cools less efficiently than metal-rich gas because metals are much more efficient coolants than hydrogen. 
In such a metal-poor and high-temperature environment, protostellar cores \tcrb{exhibit} large Jeans masses, consequently evolving into massive stars.  
Performing cosmological zoom-in simulations, \citet{Hirano2015} obtain $\sim300$ M$_{\odot}$ stars in primordial star-forming clouds. 
\citet{Hirano2015} also report the mass distribution of first stars, which is indicative that very massive ($\gtrsim100$ M$_{\odot}$) stars are born in the metal-free environment. 

Metal-poor (and thus young) galaxies potentially undergo chemical evolutions largely affected by massive stars.
Especially, iron-to-oxygen (Fe/O) abundance ratios strongly depend on galaxy ages and initial mass functions (IMFs) because stars with different masses undergo different types of supernova (SN) explosions ejecting different amount of iron and oxygen \tcrg{(Section \ref{subsec:sn})}.
\tcrb{If we calculate ejecta from Type Ia SNe and core-collapse SNe (CCSNe) only, the Fe/O ratio monotonically increases with age (and thus can serve as a cosmic clock; \citealt{Xing2019}).
%However, hypernovae with high explosion energies and/or low mass cuts (BrHNe; \citealt{Umeda2008}) and pair-instability SNe (PISNe; \citealt{Takahashi2018}) can also produce high Fe/O gas before the appearance of Type Ia SNe.
\tcrg{Before the appearance of Type Ia SNe, however, high Fe/O gas can also be produced by hypernovae (HNe) and pair-instability SNe (PISNe), whose progenitor stars are more massive than $\sim30$ and 200 M$_{\odot}$, respectively.}
Extremely-young metal-poor galaxies may have high Fe/O ratios because both HNe and PISNe tend to be produced in metal-poor environments.
Now we should evaluate the HN or PISN contributions to Fe/O in the early galaxy formation phase to test whether Fe/O can truly act as a cosmic clock in primordial galaxies.}

\begin{figure}[t]
    \centering
    \includegraphics[width=8.0cm]{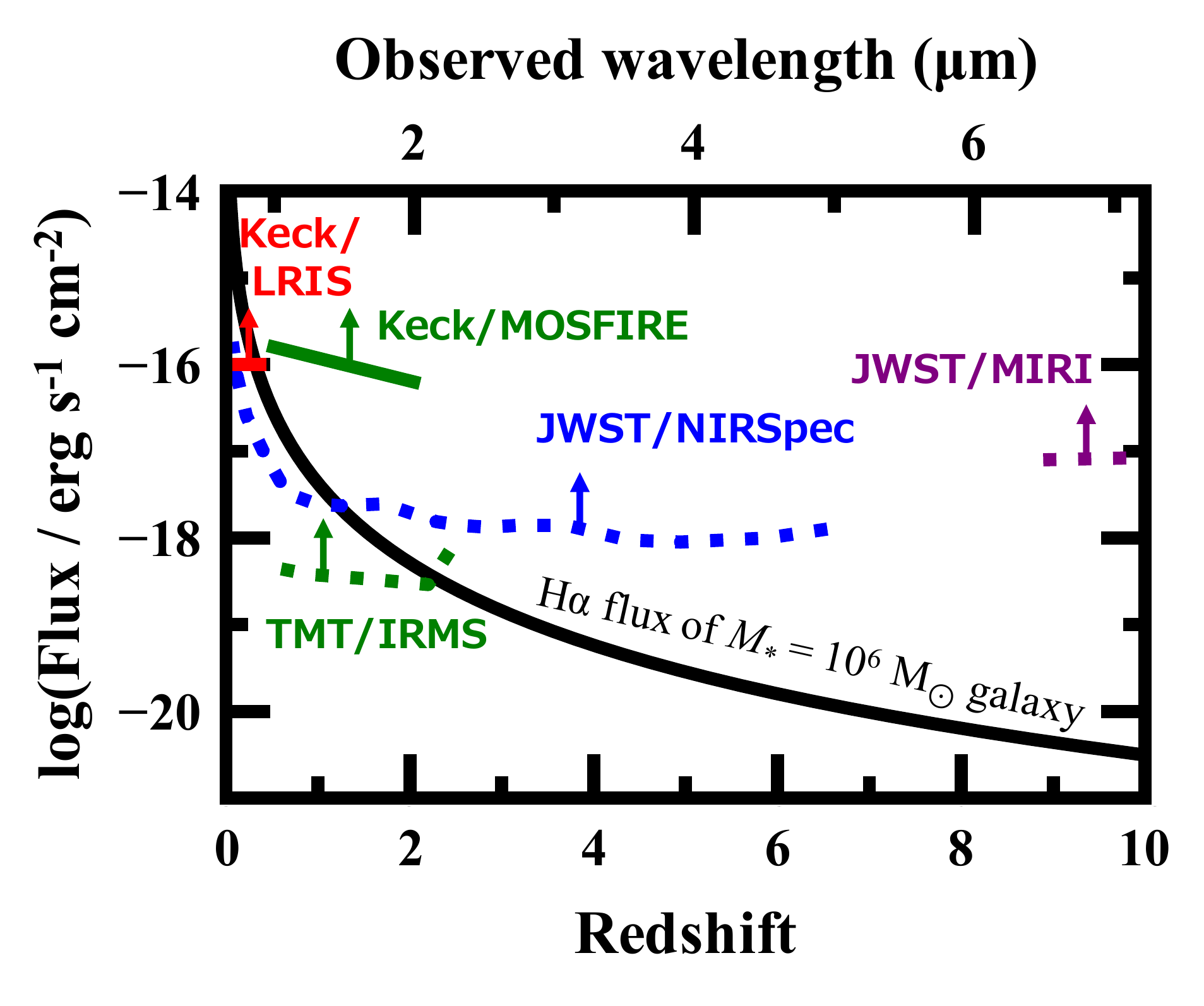}
    \caption{Expected H$\alpha$ flux of young galaxies with $M_{*}=10^{6}$ M$_{\odot}$ and \tcra{EW$_{0}$(H$\alpha)=3700$ \AA}\ at a given redshift (black solid curve). The red and green solid lines indicate the limiting fluxes of the current-best instruments of Keck/LRIS and Keck/MOSFIRE, respectively. The green, blue, and purple dotted lines represent the limiting fluxes of the near-future instruments of \tcrb{TMT/IRMS}, JWST/NIRSpec, and JWST/MIRI, respectively. \tcrb{These limiting fluxes are calculated under the assumption that a point source has an emission line that should be detected at a signal-to-noise ratio of 10 in an exposure time of 10000 second. The limiting fluxes of the instruments of TMT and JWST are derived from \citet{Hees2015} and \citet{Gardner2006}, respectively.}}
    \label{fig:obslimit}
\end{figure}

\tcrb{We can verify whether HNe and/or PISNe play important roles in the Fe/O evolution of galaxies by observing extremely young ($\lesssim10$ Myr) galaxies.}
However, such galaxies are intrinsically too faint because of their low stellar masses. 
Assuming $M_{*}=10^{6}$ M$_{\odot}$ \citep{Wise2012a} and the large rest-frame equivalent width of \tcra{EW$_{0}$(H$\alpha)=3700$ \AA\ under the assumptions of the age of 1 Myr, metal free, and constant star formation} \citep{Inoue2011}, we derive the H$\alpha$ flux of extremely-young low-mass galaxy at each redshift. 
As illustrated in Figure \ref{fig:obslimit}, the expected H$\alpha$ fluxes of the young low-mass galaxies (black) is $>1$ dex smaller than the limiting fluxes of the current-best instruments of Keck/LRIS and Keck/MOSFIRE within the redshift range of $z\gtrsim0.5$.
\tcrb{Even using the forthcoming Thirty Meter Telescope (TMT) and James Webb Space Telescope (JWST), we can detect H$\alpha$ of the young low-mass galaxies only at $z\lesssim2$.}
This means that it is difficult to detect young galaxies having $M_{*}\lesssim10^6\ {\rm M}_{\odot}$ \tcrb{at $z\gtrsim2$} even with the forthcoming large telescopes \tcra{without gravitational lensing (e.g., \citealt{Kikuchihara2020})}. 

\tcrb{Complementing} these high-$z$ galaxy observations, various studies actively investigate local young dwarf galaxies \citep[e.g.,][]{Berg2019}.
Although characteristics and formation processes of local young galaxies would be different from high-$z$ galaxies (\citealt{Isobe2020}; hereafter Paper III), local young galaxies are useful not only for studying galaxies at the early-formation stage, but also for understanding local young galaxies themselves, as living fossils of forming galaxies in the universe today.
Although such low-mass and metal-poor galaxies become rarer toward lower redshifts \citep{Behroozi2013,Morales-Luis2011}, recent studies show the presence of extremely metal-poor galaxies (EMPGs; defined as galaxies with $\leq12+\log(\rm O/H)=7.69=10$\% (O/H)$_{\odot}$) in the local universe such as SBS0335$-$052 \citep{Izotov2009}, AGC198691 \citep{Hirschauer2016}, J1234+3901 \citep{Izotov2019}, Little Cub \citep{Hsyu2017}, DDO68 \citep{Pustilnik2005}, and IZw18 \citep{Izotov1998}.
\tcrg{A blind H\,{\sc i} survey (ALFALFA) has identified Leo P with $\sim3$\% (O/H)$_\odot$ \citep{Skillman2013}.}
Significant \tcrb{progress has} been made recently with EMPG spectroscopic \tcrg{and photometric} samples of Sloan Digital Sky Survey (SDSS). 
\citet{Almeida2016} have found 196 EMPGs from the SDSS \tcrg{spectroscopic} data, and \citet{Izotov2018} have identified J0811+4730 with a metallicity down to 2\% (O/H)$_\odot$.
\tcrg{The SDSS imaging data have largely contributed to identify many EMPGs \citep[e.g.,][]{James2015,James2017,Hsyu2018,Senchyna2019}.}

However, such previous studies based on SDSS are not ideal for pinpointing low-mass (and thus faint) EMPGs due to their shallow data. 
\citeauthor{Kojima2020a} (\citeyear{Kojima2020a}; hereafter Paper I) have launched a project entitled \tcra{``}Extremely Metal-Poor Representatives Explored by the Subaru Survey (EMPRESS)'' with Subaru/Hyper Suprime-Cam (HSC) optical wide (500 deg$^2$) and deep (\tcra{5$\sigma$ limiting magnitude of $i_{\rm lim}=26$}) images that are about 100 times deeper than those of SDSS \citep{Aihara2019}. 
EMPRESS has identified J1631+4426\tcra{, whose} metallicity is 1.6\% (O/H)$_{\odot}$. 
\tcra{J1631+4426} shows the lowest metallicity reported so far, with a low stellar mass of $\sim10^{6}$ ${\rm M}_{\odot}$. 

\tcrb{While J1631+4426 (Paper I) and J0811+4730 \citep{Izotov2018} have extremely low-metallicities of $\sim2$\% (O/H)$_{\odot}$, \citeauthor{Kojima2021} (\citeyear{Kojima2021}; hereafter Paper II) have reported that the 2 EMPGs show high Fe/O ratios of $\sim(\rm Fe/O)_{\odot}$.
Paper II concludes that the 2 EMPGs are too young to be affected by chemical enrichment of low- and intermediate-mass stars because the EMPGs have low N/O ratios.
Alternatively, Paper II suggests that super-massive stars\footnote{\tcrc{More precisely, Paper II predicts super-massive stars beyond 300 M$_{\odot}$. However, such massive stars are not likely to contribute to the Fe/O enhancements of EMPGs unless the stars rotate very fast \citep{Shibata2002}.}} may contribute to the Fe/O enhancements, while the contribution has not been evaluated quantitatively.}

This paper is the fourth paper of EMPRESS, reporting spectroscopic follow-up observations for the remaining EMPG candidates with Keck Telescope. 
\tcra{We also derive chemical properties of the EMPGs including $12+\log(\rm O/H)$ and Fe/O to discuss chemical enrichment of galaxies in the early formation phase.}
We present the EMPG sample in Section \ref{sec:sample}. 
In Section \ref{sec:obs} we explain our optical spectroscopy and data reductions. 
We explain our data analysis in Section \ref{sec:analysis}.
We report and discuss chemical properties of EMPGs in Section \ref{sec:result}.
\tcrb{We discuss further the origin of the Fe/O enhancements of the EMPGs in Section \ref{sec:feo}.}
Section \ref{sec:sum} summarizes our findings. 
Throughout this paper, magnitudes are in the AB system \citep{Oke1983}, and we assume a standard $\Lambda$CDM cosmology with parameters of ($\Omega_{\rm m}$, $\Omega_{\rm \Lambda}$, $H_{0}$) = (0.3, 0.7, 70 km ${\rm s}^{-1}$ ${\rm Mpc}^{-1}$). 
The definition of solar metallicity ${\rm Z}_{\odot}$ is given by $12+\log(\rm O/H)=8.69$ \citep{Asplund2021}. Solar abundance ratios of log(Ne/O), log(Ar/O), log(N/O), log(Fe/O) are \tcra{$-0.63$, $-2.31$, $-0.86$, and $-1.23$}, respectively \citep{Asplund2021}.

\section{Sample} \label{sec:sample}
We use a photometric sample of EMPG candidates selected by Paper I. 
The Paper-I photometric sample consists of EMPG candidates identified \tcrb{from} the data of HSC and SDSS, which \tcrb{we refer to as} HSC EMPG candidates and SDSS EMPG candidates, respectively. In this paper, we do not use SDSS EMPG candidates, because the SDSS EMPG candidates include \tcra{more contaminants than} the HSC EMPG candidates (Paper I). 
The catalog of the HSC EMPG candidates is developed with the HSC-SSP S17A and S18A data \citep{Aihara2019}\tcra{, which} are wide and deep enough to search for rare and faint EMPGs. The HSC EMPG candidates are selected from $\sim46$ million sources whose photometric measurements are brighter than 5$\sigma$ limiting magnitudes in all of the 4 broadbands, $g<26.5$, $r<26.0$, $i<25.8$, and $z<25.2$ mag \citep{Ono2018}\tcrb{, which correspond to absolute magnitudes at $z=0.03$ of $M_{g}<-9.1$, $M_{r}<-9.6$, $M_{i}<-9.8$, and $M_{z}<-10.4$ mag, respectively}. The catalog consisting of these sources is referred to as the HSC source catalog.

With the HSC source catalog, Paper I isolates EMPGs from contaminants such as other types of galaxies, Galactic stars, and \tcra{quasi-stellar objects (QSOs)}. Paper I aims to find galaxies at $z\leq0.03$ 
with ${\rm EW}_{0}({\rm H}\alpha)>800$ \AA\ and $12+\log{(\rm O/H)}=6.69$--7.69. Because it is difficult to distinguish EMPGs from the contaminants on 2-color diagrams such as $r-i$ vs. $g-r$, Paper I constructs a machine-learning classifier based on a deep neural network (DNN) with a training data set. The training data set is composed of mock photometric measurements for model spectra of EMPGs and the contaminants \tcra{of other types of galaxies, Galactic stars, and QSOs}.
The DNN allows us to isolate EMPGs from the contaminants with non-linear boundaries in the multi-dimensional color space. Paper I finally obtains 27 HSC EMPG candidates from the HSC source catalog. Paper I conducts spectroscopic follow-up observations with Magellan/LDSS-3, Magellan/MagE, Keck/DEIMOS, and Subaru/FOCAS for 4 out of the 27 HSC EMPG candidates, and confirm that all of the 4 HSC EMPG candidates are truly emission-line galaxies with the low metallicity of $12+\log{(\rm O/H)}=6.90$--8.27 (i.e., 1.6--38\% ${\rm Z}_{\odot}$). 
\tcra{Paper I finds that 2 out of the 4 HSC EMPG candidates meet the EMPG criterion of $12+\log{(\rm O/H)}<7.69$ (i.e., $<10\%$ ${\rm Z}_{\odot}$).}
We refer to the 2 EMPGs as HSC spectroscopic EMPGs. 
There remain 23 ($=27-4$) HSC EMPG candidates that are not spectroscopically confirmed in Paper I. 
We refer to the 23 candidates as HSC photometric EMPGs. 

\section{Observations and Data Reduction} \label{sec:obs}
\subsection{Spectroscopic Follow-up Observations with Keck/LRIS} \label{subsec:lris}

\renewcommand{\arraystretch}{0.7}
\begin{table}[t]
    \begin{center}
    \caption{Coordinates of the HSC EMPG candidates}
    \label{tab:coord}
    \begin{tabular}{cccc} \hline \hline
        \# & ID & R.A. & Dec. \\
        & & hh:mm:ss & dd:mm:ss \\
        (1) & (2) & (3) & (4) \\ \hline
        1 & J0156$-$0421 & 01:56:51.6 & $-$04:21:25.2 \\
        2 & J0159$-$0622 & 01:59:43.8 & $-$06:22:32.8 \\
        3 & J0210$-$0124 & 02:10:12.0 & $-$01:24:51.1 \\
        4 & J0214$-$0243 & 02:14:24.3 & $-$02:43:54.4 \\
        5 & J0226$-$0517 & 02:26:57.6 & $-$05:17:47.3 \\
        6 & J0232$-$0248 & 02:32:13.3 & $-$02:48:19.3 \\
        7 & J1608$+$4337 & 16:08:11.0 & +43:37:53.4 \\
        8 & J2236$+$0444 & 22:36:12.4 & +04:44:22.3 \\
        9 & J2321$+$0125 & 23:21:52.2 & +01:25:55.0 \\
        10 & J2355$+$0200 & 23:55:30.1 & +02:00:16.0 \\
        11 & J0228$-$0256 & 02:28:36.3 & $-$02:56:45.7 \\
        12 & J2221$-$0015 & 22:21:26.1 & $-$00:15:49.5 \\
        13 & J2319$+$0136 & 23:19:33.6 & +01:36:50.2 \\ \hline
    \end{tabular}
    \end{center}
    \tablecomments{(1): Number. (2): ID. (3): Right ascension in J2000. (4): Declination in J2000. }
\end{table}

In this paper, we report spectroscopic observations with \tcrb{the} Low Resolution Imaging Spectrograph (LRIS\tcra{; \citealt{Oke1995}}). 
LRIS is an imaging spectrometer installed at the Cassegrain focus of Keck Telescope\tcra{, whose} aperture area of is equivalent to that with a circular aperture of 9.96 m in diameter. 
LRIS has both blue and red channels that roughly cover wavelength ranges of 3000--6000 and 6000--10000 \AA, respectively. 
LRIS can perform long-slit spectroscopy or multi-object spectroscopy (MOS). 

We conducted spectroscopy with Keck/LRIS (PI: T. Kojima) for 13 out of the 23 HSC photometric EMPGs. 
\tcrb{The 13 targets were all the HSC photometric EMPGs} observable at the observing night of 2019 August 31. 
Coordinates of the 13 HSC photometric EMPGs are listed in Table \ref{tab:coord}.
We utilized the MOS mode and long-slit mode for 7 and 6 candidates, respectively. 
The slit widths were 1.5 arcsec for the all targets. 
We used the 600 lines mm$^{-1}$ grism blazed at 4000 \AA\ on the blue channel and the 600 lines mm$^{-1}$ grating blazed at 7500 \AA\ on the red channel. 
The LRIS spectroscopy of the blue and red channels covered the wavelength ranges of $\lambda\sim3000$--5500 and 6000--9000 \AA\ with the spectral resolutions of $\sim4$ and 5 \AA\ in FWHM, respectively. 
We also observed standard stars of \tcrb{a} DOp-type star Feige 110 (RA=23:19:58.4, Dec.=$-$05:09:56 in J2000), \tcrb{a} B2III-type star BD+40 4032 (RA=20:06:40.0, Dec.=+41:06:15 in J2000), and \tcrb{a} B6V-type star Feige 25 (RA=02:36:00.0, Dec.=+05:15:17 in J2000). 
The sky was clear during the observations with seeing sizes of 0.8 arcsec.
The observations are summarized in Table \ref{tab:obs}. 

\begin{figure*}[t]
    \centering
    \includegraphics[width=18.0cm]{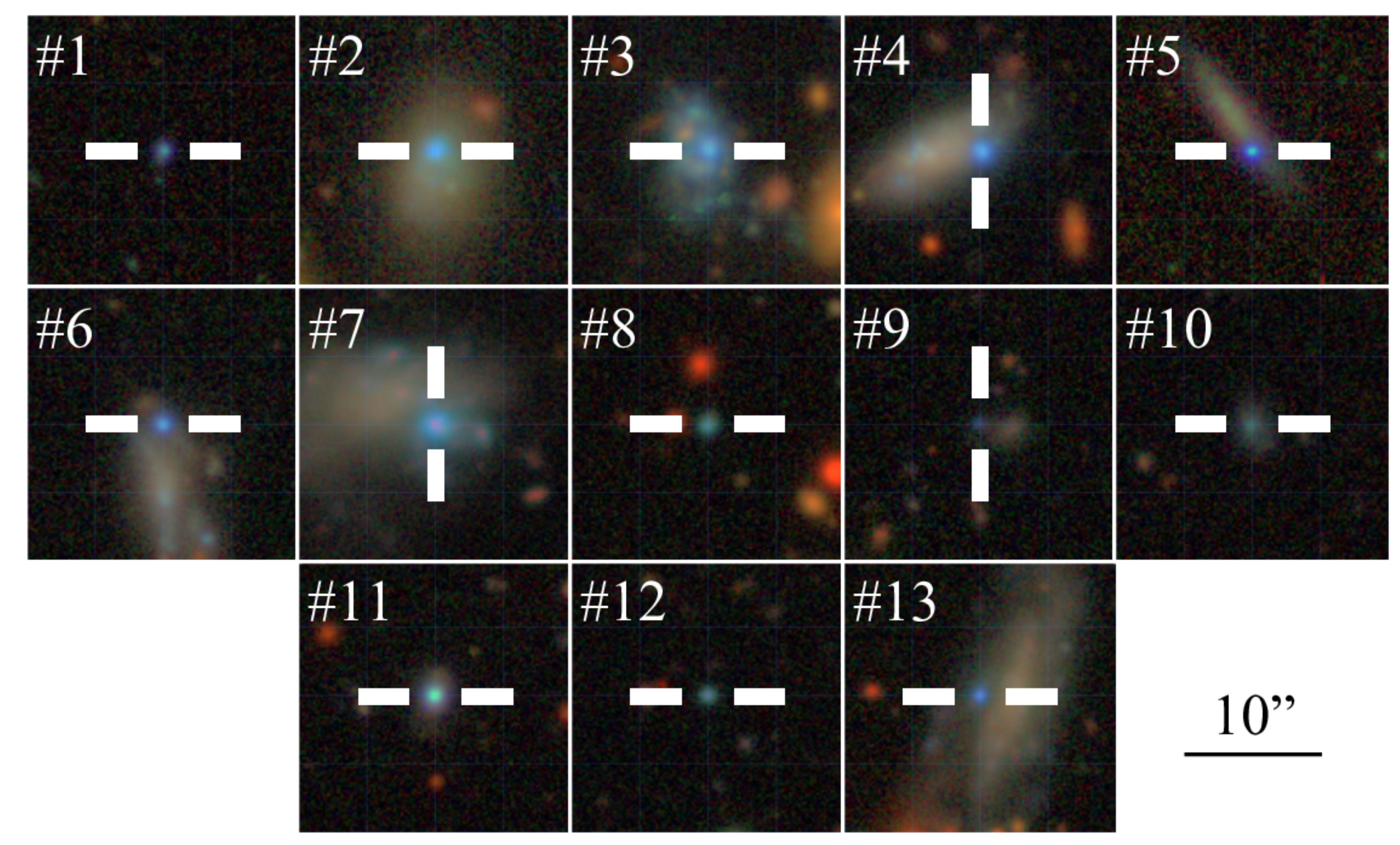}
    \caption{HSC $gri$-composite images of the 13 HSC photometric EMPGs that we observed with LRIS. The HSC $g$, $r$, and $i$-bands correspond to blue, green, and red colors in the figure, respectively. Each EMPG is located between the 2 white bars of each panel. The cut-out size is $20"\times20"$. The number shown at the top left corner of each panel corresponds to the number in Table \ref{tab:obs}.}
    \label{fig:img}
\end{figure*}

Figure \ref{fig:img} shows HSC images of the 13 HSC photometric EMPGs that we observed with LRIS. 
\tcra{We note that many of the HSC photometric EMPGs (\#2, 3, 4, 5, 6, 7, 9, 10, 11, and 13 in Figure \ref{fig:img}) have diffuse structures (EMPG-tails; Paper III).
We discuss the contribution of the EMPG-tails to the flux measurement of the HSC photometric EMPGs in Section \ref{subsec:flux}.}

\begin{threeparttable}[t]
    \centering
    \caption{Summary of the LRIS observations}
    \label{tab:obs}
    \begin{tabular}{cccc} \hline \hline
        \# & ID & Mode & Exposure \\
        & & & sec \\
        (1) & (2) & (3) & (4) \\ \hline
        1 & J0156$-$0421 & Long slit & 1200 \\
        2 & J0159$-$0622 & Long slit & 1200 \\
        3 & J0210$-$0124 & MOS & 1200 \\
        4 & J0214$-$0243 & MOS & 1200 \\
        5 & J0226$-$0517 & MOS & 1200 \\
        6 & J0232$-$0248 & MOS & 1200 \\
        7 & J1608$+$4337 & MOS & 1800 \\
        8 & J2236$+$0444 & Long slit & 2400 \\
        9 & J2321$+$0125 & Long slit & 3600 \\
        10 & J2355$+$0200 & MOS & 2400 \\
        11\tnote{a} & J0228$-$0256 & Long slit & 1200 \\
        12\tnote{b} & J2221$-$0015 & Long slit & 2400 \\
        13\tnote{b} & J2319$+$0136 & Long slit & 600 \\ \hline
    \end{tabular}
    \tablecomments{(1): Number. (2): ID. (3): Observing mode (Section \ref{subsec:lris}). (4): Exposure time. }
    \begin{tablenotes}
        \item[a] \tcra{Contaminant: metal-rich galaxy (see Section \ref{subsec:reduc})}
        \item[b] \tcra{Contaminant: no emission line (see Section \ref{subsec:reduc})}
    \end{tablenotes}
\end{threeparttable}

\subsection{Data Reduction} \label{subsec:reduc}
To reduce and calibrate the data taken with LRIS, we use the {\sc iraf} package. 
The reduction and calibration processes include the bias subtraction, flat fielding, cosmic ray cleaning, sky subtraction, wavelength calibration, one-dimensional (1D) spectrum extraction, flux calibration, atmospheric-absorption correction, and Galactic-reddening correction. 
The 1D spectra are derived from apertures centered on the blue compact component of the HSC photomeric EMPGs. 
We use the standard star Feige 110 for the flux calibration. 
The wavelengths are calibrated with the HgNeArCdZnKrXe lamp. 
We correct the atmospheric absorption with the extinction curve at Mauna Kea Observatories \tcra{\citep{Beland1988}}. 
The Galactic-reddening value for each target is drawn from the NASA/IPAC Infrared Science Archive (IRSA)\footnote{\url{https://irsa.ipac.caltech.edu/applications/DUST/}} based on the \citet{Schlafly2011} estimates. 

\begin{figure*}[t]
    \centering
    \includegraphics[width=18.0cm]{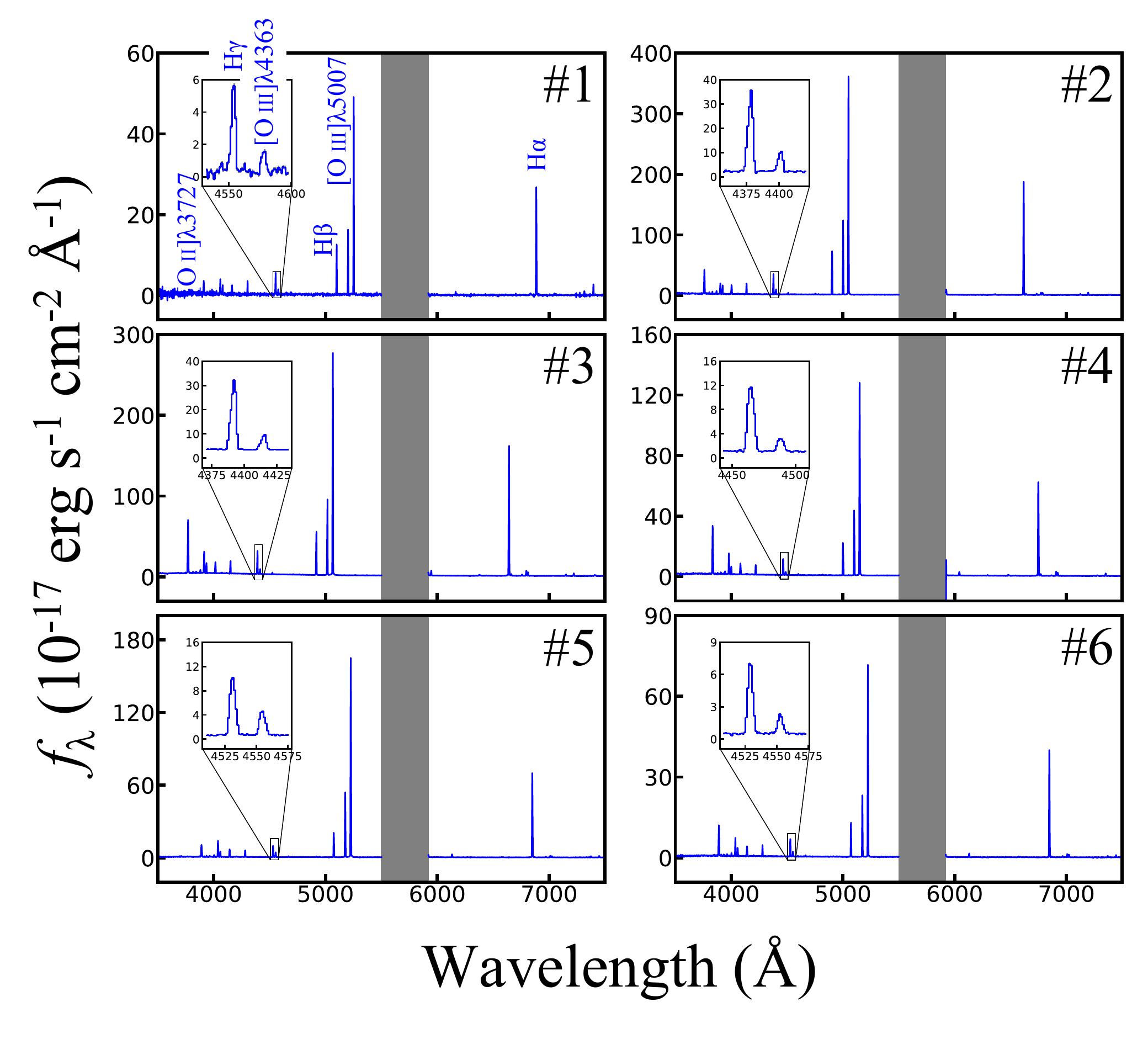}
    \caption{Reduced spectra of 6 out of the 11 HSC photometric EMPGs. The gray shaded regions indicate the gaps between the LRIS blue and red channels. The number shown at the top right corner of each panel corresponds to the number in Table \ref{tab:flux}. \tcrb{The inset panel at the top left corner of each panel illustrates an enlarged view of the spectrum around H$\gamma$ and [O\,{\sc iii}]$\lambda$4363. The gray lines indicate errors of the spectra.}}
    \label{fig:spec1}
\end{figure*}

\begin{figure*}[t]
    \centering
    \includegraphics[width=18.0cm]{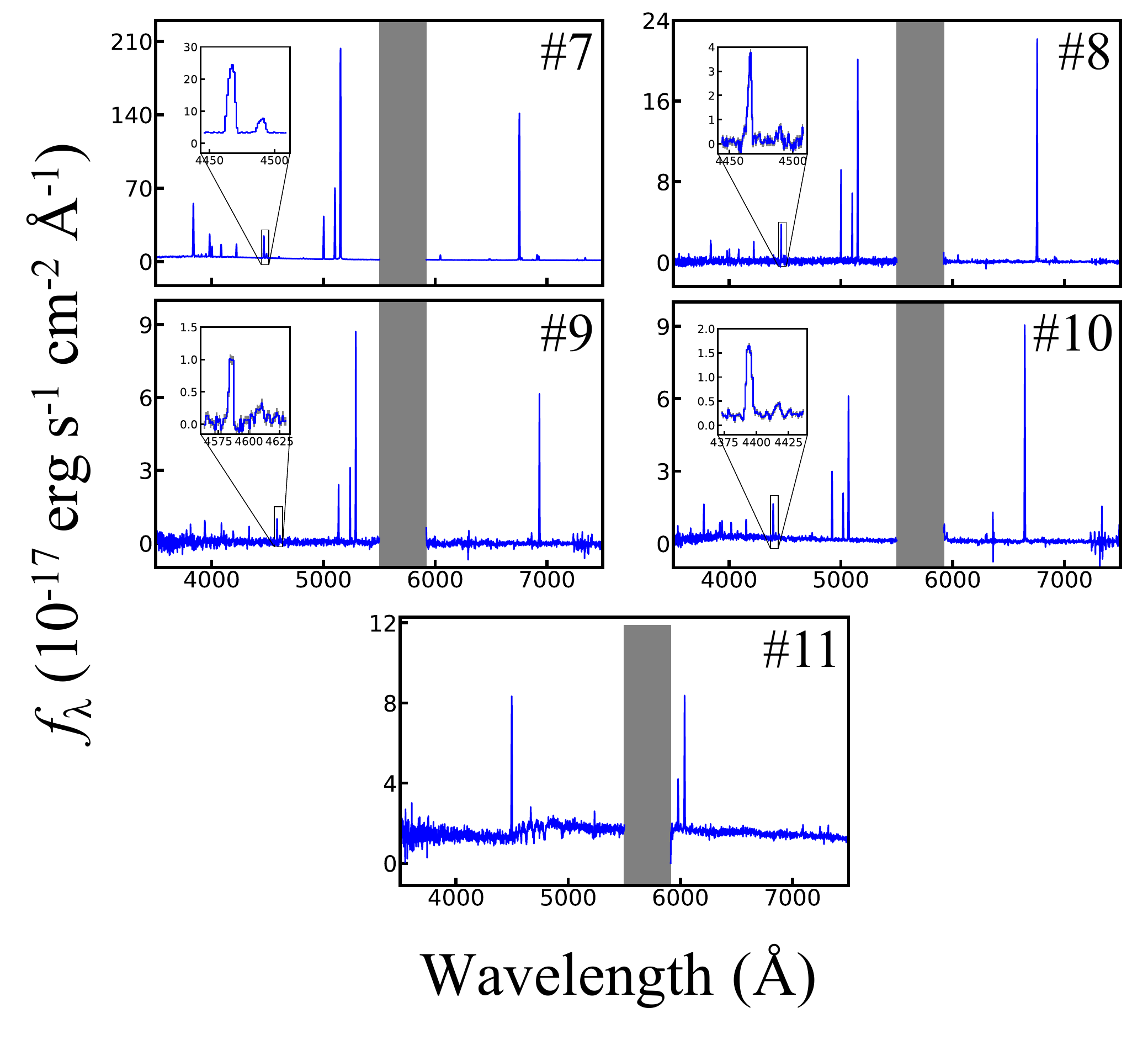}
    \caption{Reduced spectra of the remaining 5 of the 11 HSC photometric EMPGs. The symbols are the same as \tcrd{those} in Figure \ref{fig:spec1}. J0228$-$0256 (\#11) is a \tcrc{metal-rich galaxy at $z=0.21$ (Section \ref{subsec:reduc})}. The other 10 HSC photometric EMPGs are EMPG candidates at $z=0.009$--0.056.}
    \label{fig:spec2}
\end{figure*}

We detect emission lines from 11 out of the 13 HSC photometric EMPGs. 
The \tcrb{remaining} 2 HSC photometric EMPGs \tcra{(J2221$-$0015 and J2319+0136; \#12 and 13)} with no emission line detected are probably contaminants, because the H$\alpha$ fluxes estimated from $r$-band excesses are $4\times10^{-15}$--$1\times10^{-14}$ erg s$^{-1}$ cm$^{-2}$\tcra{, which} should be detectable with the 10-minute exposure of LRIS. 
Figures \ref{fig:spec1} and \ref{fig:spec2} presents the reduced spectra of the 11 HSC photometric EMPGs.
\tcra{One out of the 11 HSC photometric EMPGs (J0228$-$0256; \#11 in Figure \ref{fig:spec2}) is located at $z=0.21$, which is out of the redshift range where Paper I aims to select EMPGs (Section \ref{sec:sample}). 
Because \tcrb{\#11} shows a high [O\,{\sc ii}]$\lambda\lambda$3727,3729/[O\,{\sc iii}]$\lambda$5007 ratio and a strong Balmer break, \tcrb{\#11} is likely to be a metal-rich galaxy.} The remaining 10 HSC photometric EMPGs appear to be located at the redshifts of $z=0.009$--0.056, where Paper I aims to select EMPGs. 
Hereafter we refer to these 10 HSC photometric EMPGs as LRIS EMPG candidates.
\tcrg{The inset panels of Figures \ref{fig:spec1} and \ref{fig:spec2} indicate that many of the LRIS EMPG candidates have emission lines with blue wings, suggestive of the outflow (Xu et al. in prep.; cf. Section \ref{subsec:ratio}).}

\section{Analysis} \label{sec:analysis}
\subsection{Flux Measurement} \label{subsec:flux}
We measure central wavelengths, emission-line fluxes, \tcrb{and continua} of the LRIS EMPG candidates (Section \ref{subsec:reduc}) with best-fit Gaussian \tcrb{(+ constant)} profiles using the {\tt scipy.optimize} package. 
We also estimate flux errors containing read-out noise and photon noise of sky and object emissions. 
\tcra{None of the LRIS EMPG candidates show broad Balmer lines or high-ionization lines such as [Fe\,{\sc vii}]$\lambda$6087, which suggests that the radiation from active galactic nuclei (AGNs) is not dominant in any LRIS EMPG candidate.}
We detect faint [O\,{\sc iii}]$\lambda$4363 lines from all the 10 LRIS EMPG candidates. 
\tcrg{Because none of the LRIS EMPG candidates show [Fe\,{\sc ii}]$\lambda$4288, whose flux is larger than that of [Fe\,{\sc ii}]$\lambda$4359\footnote{PyNeb provides $[{\rm Fe}\,\textsc{ii}]\lambda4359=0.73\times[{\rm Fe}\,\textsc{ii}]\lambda4288$ under the assumptions of $T_{\rm e}=20000$ K and $n_{\rm e}=100$ cm$^{-3}$.}, the contamination of [O\,{\sc iii}]$\lambda$4363 from [Fe\,{\sc ii}]$\lambda$4359 is negligible.}
We also obtain very faint [Fe\,{\sc iii}]$\lambda$4658 lines from 2 out of the 10 LRIS EMPG candidates \tcrb{(\#2 and 7)}, which enables us to calculate Fe/O abundance ratios.  
\tcra{This flux measurement probably represents an average of the whole galaxy because the size of the typical HSC EMPG is $\sim0.3$ arcsec ($\sim200$ pc; Paper III), which is smaller than the seeing size of 0.8 arcsec and the slit width of 1.5 arcsec.} 
Redshifts are derived from the \tcra{ratios} between the observed central wavelengths and the rest-frame wavelengths in the air of H$\beta$ lines. 

Color excesses $E(B-V)$ are derived from the Balmer decrement under the assumptions of the dust \tcra{attenuation} curve of \citet{Calzetti2000}\footnote{\tcra{A choice of attenuation curves does not change our results significantly because the Balmer decrements of the LRIS EMPG candidates are comparable to those under the assumption of the case B recombination (i.e., dust-poor). }} 
and the case B recombination.
\tcrg{We calculate $E(B-V)$, electron temperature $T_{\rm e}$, and electron density $n_{\rm e}$ iteratively so that all these properties are consistent with each other (see also Section \ref{subsec:chem}).
We obtain intrinsic values of Balmer emission-line ratios using PyNeb (\citealt{Luridiana2015}; v1.1.15).}
\tcrb{We derive $\chi^{2}$ of $E(B-V)$ from each Balmer emission-line ratio containing H$\alpha$, H$\beta$, H$\gamma$, and H$\delta$.
Then, we find the best $E(B-V)$ values, which give the least $\chi^{2}$.
We also obtain $\pm68$\% confidence intervals of $E(B-V)$ based on $\chi^{2}$.}
From all the LRIS EMPG candidates, we detect H$\gamma$ \tcrb{and H$\delta$} lines with ${\rm S/N}>13$ \tcrb{and ${\rm S/N}>5$, respectively.}
We also confirm that none of the \tcrb{H$\delta$, H$\gamma$, H$\beta$, and H$\alpha$} lines show significant stellar absorption.
Using the $E(B-V)$ and the \citet{Calzetti2000} \tcra{attenuation} curve, we derive dust-corrected fluxes. 
We note that the dust-corrected fluxes are not very different from the observed ones, because most of the LRIS EMPG candidates show $E(B-V)\sim0$ (i.e., dust-poor). 
The dust-corrected fluxes are summarized in Table \ref{tab:flux}. 
\tcra{Again, we} note that we detect the weak [Fe\,{\sc iii}]$\lambda$4658 lines from 2 out of the 10 LRIS EMPG candidates, which are used to derive the Fe abundance. 
Other fundamental properties such as the redshift, rest-frame equivalent width of EW$_{0}$(H$\beta$), and $E(B-V)$ are listed in Table \ref{tab:prop}. 
%\tcra{Checking 2D spectrum, we note that emission-line and continuum fluxes of the EMPG-tails \tcrb{(See Section \ref{subsec:lris})} can contaminate those of the LRIS EMPG candidates \tcrb{(especially \#2--7)} by at most $\sim10$ and $\sim50$\%, respectively.
%This indicates that emission lines of the LRIS EMPG candidates are not contaminated significantly, while EW$_{0}$(H$\beta$) values are potentially underestimated.}
Checking 2D spectra, \tcrg{we find that emission lines of EMPG-tails can contaminate those of EMPGs by at most 10\%.
We then add the uncertainties to lower errors of fluxes of EMPGs with EMPG-tails.
On the other hand, stellar continua of EMPG-tails potentially contaminate those of EMPGs by at most 50\%, which implies that EW$_{0}$(H$\beta$) of EMPGs would be underestimated.
We calculate errors of EW$_{0}$(H$\beta$) including the uncertainties of stellar continua as well as those of H$\beta$ fluxes.}

\subsection{Chemical Property} \label{subsec:chem}
\begin{table*}[t]
    \begin{center}
    \caption{\tcrg{Atomic data}}
    \label{tab:atom}
    \begin{tabular}{cccc} \hline \hline
		Ion & Emission process & Transition probability & Collision Strength \\
		(1) & (2) & (3) & (4) \\ \hline
		H$^{0}$ & Re & \citet{Storey1995} & --- \\
		O$^{+}$ & CE & \citet{FroeseFischer2004} & \citet{Kisielius2009} \\
		O$^{2+}$ & CE & \citet{FroeseFischer2004} & \citet{Storey2014} \\
		Ne$^{2+}$ & CE & \citet{FroeseFischer2004} & \citet{McLaughlin2000} \\
		Ar$^{2+}$ & CE & \citet{MunozBurgos2009} & \citet{MunozBurgos2009} \\
		N$^{+}$ & CE & \citet{FroeseFischer2004} & \citet{Tayal2011} \\
		Fe$^{2+}$ & CE & \citet{Quinet1996,Johansson2000} & \citet{Zhang1996} \\ \hline
    \end{tabular}
    \end{center}
    \tablecomments{(1): Ion. (2): Emission process. Re and CE represent recombination and collisional excitations, repsectively. (3): Reference of the transition probability of each ion. (4): Reference of the collision strength of each ion. We note that these atomic data and temperature relationships are the latest ones applied in \citet{Berg2015} and Paper II.}
\end{table*}
All of the LRIS EMPG candidates \tcrb{have [O\,{\sc iii}]$\lambda$4363 detections}, which \tcra{allows} us to derive metallicities and other element abundances with the direct-$T_{\rm e}$ method \citep[\tcrg{e.g.,}][]{Izotov2006} as described below. 

The electron temperature $T_{\rm e}$ is calculated from two collisional excitation lines of the same ion such as O$^{2+}$, because the collisional excitation rate is determined by $T_{\rm e}$.
\tcrg{Using the PyNeb package {\tt getCrossTemDen} with the latest atomic data and temperature relationship listed in Table \ref{tab:atom}, we derive $T_{\rm e}$ of O$^{2+}$ ($T_{\rm e}$(O\,{\sc iii})) and $n_{\rm e}$ from emission-line ratios of [O\,{\sc iii}]$\lambda$4363/[O\,{\sc iii}]$\lambda\lambda$4959,5007 and [S\,{\sc ii}]$\lambda$6731/[S\,{\sc ii}]$\lambda$6716, respectively. If [S\,{\sc ii}]$\lambda$6716 is not available, we calculate $T_{\rm e}$ with a fixed $n_{\rm e}$ of 100 cm$^{-3}$, which is roughly consistent with those of EMPGs (e.g., Paper I).}
\if0
As an example, we explain the common method to derive $T_{\rm e}$ of O$^{2+}$ ions using the following equation \citep{Osterbrock1989}:
\begin{equation}
    \frac{F([{\textsc{O\,iii}}]\lambda\lambda4959,5007)}{F([{\textsc{O\,iii}}]\lambda4363)}=\frac{7.90\exp(3.29\times10^{4}/T_{\rm e})}{1+4.5\times10^{-4}n_{\rm e}/T^{1/2}_{\rm e}},
    \label{equ:te}
\end{equation}
where $n_{\rm e}$ represents the electron density. 
The [O\,{\sc iii}]$\lambda$4363 and [O\,{\sc iii}]$\lambda\lambda$4959,5007 lines are emitted by the downward transitions of $^{1}{\rm S}_{0}\rightarrow$ $^{1}{\rm D}_{2}$ and $^{1}{\rm D}_{2}\rightarrow$ $^{3}{\rm P}_{1,2}$, respectively.
Because a typical \tcra{H\,{\sc ii} region cloud} has a $n_{\rm e}$ value \tcra{much lower than a critical density of $6.4\times10^{5}$ cm$^{-3}$ for the \tcrb{transition of $^{1}{\rm S}_{0}\rightarrow$ $^{1}{\rm D}_{2}$},} $T_{\rm e}$ is mainly determined by the ratio $F([{\textsc{O\,iii}}]\lambda\lambda4959,5007)/F([{\textsc{O\,iii}}]\lambda4363)$.
The ratio $F([{\textsc{O\,iii}}]\lambda\lambda4959,5007)/F([{\textsc{O\,iii}}]\lambda4363)$ decreases with $T_{\rm e}$ because both the collisional excitation rate from $^{1}{\rm D}_{2}$ to $^{1}{\rm S}_{0}$ and the collisional de-excitation rate from $^{1}{\rm D}_{2}$ to $^{3}{\rm P}_{1,2}$ \tcra{increase with $T_{\rm e}$}. 
\fi
\tcrc{We summarize results of $T_{\rm e}$ \tcrg{and $n_{\rm e}$} in Table \ref{tab:prop}.}
%We confirm that most of the LRIS EMPG candidates have $T_{\rm e}\sim20000$ K, which is consistent with \tcrd{the assumption that $T_{\rm e}=20000$ K for our $E(B-V)$ estimates} (Section \ref{subsec:flux}).
\tcrg{Again, our iterative calculations provide self-consistent values of $E(B-V)$, $T_{\rm e}$(O\,{\sc iii}), and $n_{\rm e}$ (cf. Section \ref{subsec:flux}).
We also note that} $T_{\rm e}$ is mainly determined by the ratio $F([{\textsc{O\,iii}}]\lambda4363)/F([{\textsc{O\,iii}}]\lambda\lambda4959,5007)$ \tcrg{and almost independent of $n_{\rm e}$} because \tcrg{the derived $n_{\rm e}$ values are} much lower than a critical density of $6.4\times10^{5}$ cm$^{-3}$ for the \tcrg{[O\,{\sc iii}]$\lambda$4363} transition of $^{1}{\rm S}_{0}\rightarrow$ $^{1}{\rm D}_{2}$.

\tcrg{Using PyNeb with the latest atomic data and temperature relationships listed in Table \ref{tab:atom}, we derive ion abundance ratios of O$^{+}$/H$^{+}$ and O$^{2+}$/H$^{+}$ from emission-line ratios of [O\,{\sc ii}]$\lambda\lambda$3727,3729/H$\beta$ and [O\,{\sc iii}]$\lambda\lambda$4959,5007/H$\beta$ with electron temperatures of $T_{\rm e}$(O\,{\sc ii}) and $T_{\rm e}$(O\,{\sc iii}), respectively, and the derived $n_{\rm e}$.
We calculate $T_{\rm e}$(O\,{\sc ii}) using an empirical relation of
\begin{equation}
    T_{\rm e}(\textsc{O\,ii})=0.7T_{\rm e}(\textsc{O\,iii})+3000
    \label{equ:teo2}
\end{equation}
\citep{Garnett1992}. 
Just adding O$^{+}$/H$^{+}$ to O$^{2+}$/H$^{+}$, we finally obtain $12+\log(\rm O/H)$.}
\tcra{We note that the neutral oxygen is negligible in H\,{\sc ii} regions because \tcrb{the ionization potential of neutral oxygen atoms is 13.6 eV, the same as that of neutral hydrogen atoms}.}
\tcrb{We also ignore O$^{3+}$ and higher-order oxygen ions for consistency with previous works (e.g., \citealt{Izotov2006}; Paper I)\footnote{\tcrb{O$^{3+}$ and higher-order oxygen ions are usually ignored because they have ionization potentials of $>55$ eV, which cannot be generated by UV radiation of typical stars.
However, super-massive or metal-free stars can efficiently produce high-energy photons above 55 eV (\citealt{Vink2018} and \citealt{Tumlinson2000}, respectively). If such stars exist in EMPGs (e.g., Paper II), the higher-order oxygen ions may not be negligible.}}.}

Using \tcrg{the latest atomic data and temperature relationships listed in Table \ref{tab:atom}}, we can also derive other \tcra{gas-phase} ion abundances such as Ne$^{2+}$/H$^{+}$, Ar$^{2+}$/H$^{+}$, N$^{+}$/H$^{+}$, and Fe$^{2+}$/H$^{+}$ with optical emission lines of [Ne\,{\sc iii}]$\lambda$3869, [Ar\,{\sc iii}]$\lambda$7136, [N\,{\sc ii}]$\lambda\lambda$6548,6584, and [Fe\,{\sc iii}]$\lambda$4658, respectively. 
\tcrb{We use $T_{\rm e}({\textsc{O\,iii}})$ and $T_{\rm e}({\textsc{S\,iii}})$ to calculate high- and intermediate-ionization ion abundances of Ne$^{2+}$ and Ar$^{2+}$, respectively. 
We derive $T_{\rm e}({\textsc{S\,iii}})$ from an empirical relation of}
\begin{equation}
    T_{\rm e}({\textsc{S\,iii}})=0.83\times T_{\rm e}({\textsc{O\,iii}})+1700
    \label{equ:tes3}
\end{equation}
\tcrb{\citep{Garnett1992}.
We adopt $T_{\rm e}({\textsc{O\,ii}})$ to estimate low-ionization ion abundances of N$^{+}$ and Fe$^{2+}$.} 

\tcrg{Using the ionization correction factor (ICF) of \citet{Izotov2006}, which can be described by O$^{+}$ and O$^{2+}$ ions, we derive a total gas-phase abundance of each element from each ion abundance.}
%A total \tcra{gas-phase} abundance of each element can be derived from each ion abundance and ionization correction factor (ICF) that can be described by O$^{+}$ and O$^{2+}$ ions \citep{Izotov2006}. 
In the case of \tcrb{iron}, for example, the abundance ratio Fe/H is calculated by
\begin{equation}
    \frac{\rm Fe}{\rm H}=\frac{{\rm Fe}^{2+}}{{\rm H}^{+}}\times{\rm ICF}({\rm Fe}^{2+}).
    \label{equ:fe}
\end{equation}
It should be noted that the ICFs slightly depend on the metallicity \tcra{as follows:
\begin{eqnarray}
    {\rm ICF}({\rm Fe}^{2+})&=&0.158v+0.958+0.004/v,\ {\rm low}\ Z\\\nonumber
    &=&0.104v+0.980+0.001/v,\ {\rm intermed.}\ Z\\\nonumber
    &=&0.238v+0.931+0.004/v,\ {\rm high}\ Z,
    \label{equ:icf}
\end{eqnarray}
where $v={\rm O}^{+}/({\rm O}^{+}+{\rm O}^{2+})$. }
We adopt low-, intermediate-, and high-metallicity ICFs for galaxies with $12+\log(\rm O/H)\leq7.2$, $7.2<12+\log(\rm O/H)<8.2$, and $12+\log(\rm O/H)\ge8.2$, respectively \citep{Izotov2006}. 
\tcrg{We also estimate iron abundances using the ICF of \citet{Rodriguez2005}, which incorporates Fe$^{3+}$ abundances.
We confirm that Fe/O ratios based on \citet{Rodriguez2005} are $\sim0.2$ dex lower than those of \citet{Izotov2006} as discussed in Paper II.
We add the offsets to lower errors of Fe/O.}

In order to estimate the errors of the \tcra{gas-phase} element abundance ratios, \tcrb{we randomly fluctuate flux values based on the flux errors.
We calculate the abundance ratios 1000 times, and then obtain median values with $\pm68$\% confidence intervals of the abundance ratios.}
\tcrg{We note that our LRIS deep spectroscopy provides high S/N ratios of [O\,{\sc iii}]$\lambda$4363 especially for LRIS EMPG candidates \#1--7, which result in small errors of $12+\log(\rm O/H)$.}

We summarize the results in Table \ref{tab:ratio}. 
\tcra{Because all of the LRIS EMPG candidates are dust poor (Section \ref{subsec:flux}), the gas-phase element abundance ratios of the LRIS EMPG candidates are expected to be comparable to total element abundance ratios.}

\begin{table*}[t]
    \begin{center}
    \caption{Dust-corrected fluxes of the LRIS EMPG candidates}
    \label{tab:flux}
    \begin{tabular}{ccccccccc} \hline \hline
		\# & ID & [O\,{\sc ii}]$\lambda\lambda$3727,3729 & [Ne\,{\sc iii}]$\lambda$3869 & H$\delta$ & \tcrg{[Fe\,{\sc ii}]$\lambda$4288} & H$\gamma$ & [O\,{\sc iii}]$\lambda$4363 & [Fe\,{\sc iii}]$\lambda$4658 \\
		(1) & (2) & (3) & (4) & (5) & (6) & (7) & (8) & (9) \\ \hline
		1 & J0156$-$0421 & $38.6\pm1.5$ & $31.4\pm1.4$ & $25.1\pm1.1$ & $<4.4$ & $45.8\pm1.8$ & $11.0\pm1.2$ & $<5.3$ \\
		2 & J0159$-$0622 & $78.6\pm0.5$ & $29.6\pm0.3$ & $26.6\pm0.2$ & $<0.7$ & $47.2\pm0.4$ & $11.6\pm0.2$ & $0.8\pm0.4$ \\
		3 & J0210$-$0124 & $169.5\pm0.6$ & $50.2\pm0.3$ & $31.0\pm0.3$ & $<0.6$ & $53.4\pm0.6$ & $11.1\pm0.3$ & $<1.1$ \\
		4 & J0214$-$0243 & $170.4\pm0.7$ & $61.6\pm0.5$ & $30.2\pm0.5$ & $<1.4$ & $51.5\pm0.7$ & $10.1\pm0.5$ & $<1.8$ \\
		5 & J0226$-$0517 & $66.0\pm0.6$ & $75.1\pm0.6$ & $30.2\pm0.5$ & $<1.6$ & $50.8\pm0.8$ & $21.0\pm0.6$ & $<1.6$ \\
		6 & J0232$-$0248 & $127.9\pm1.0$ & $53.4\pm0.6$ & $31.5\pm0.7$ & $<2.2$ & $52.7\pm1.2$ & $13.5\pm0.7$ & $<2.8$ \\
		7 & J1608$+$4337 & $148.9\pm0.5$ & $53.3\pm0.3$ & $31.5\pm0.3$ & $<0.7$ & $54.2\pm0.5$ & $11.0\pm0.4$ & $1.4\pm0.5$ \\
		8 & J2236$+$0444 & $49.3\pm3.8$ & $16.3\pm2.3$ & $30.4\pm2.3$ & $<5.5$ & $41.0\pm2.0$ & $7.8\pm2.3$ & $<5.8$ \\
		9 & J2321$+$0125 & $58.4\pm5.1$ & $28.6\pm3.1$ & $23.5\pm3.1$ & $<14.5$ & $50.3\pm4.0$ & $14.5\pm6.3$ & $<16.9$ \\
		10 & J2355$+$0200 & $62.8\pm1.8$ & $17.1\pm1.4$ & $28.7\pm1.6$ & $<6.6$ & $52.9\pm2.2$ & $7.3\pm2.1$ & $<7.4$ \\ \hline \hline
		\# & He\,{\sc ii}\,$\lambda$4686 & [Ar\,{\sc iv}]$\lambda$4711 & [Ar\,{\sc iv}]$\lambda$4740 & H$\beta$ & [O\,{\sc iii}]$\lambda$4959 & [O\,{\sc iii}]$\lambda$5007 & [C\,{\sc iv}]$\lambda$5808 & He\,{\sc i}\,$\lambda$5876 \\
		(1) & (10) & (11) & (12) & (13) & (14) & (15) & (16) & (17) \\ \hline
		1 & $<5.4$ & $<4.2$ & $<5.7$ & $100.0\pm3.1$ & $133.7\pm2.9$ & $394.9\pm4.6$ & $<2.4$ & $8.9\pm0.8$ \\
		2 & $<0.9$ & $<1.9\pm0.2$ & $1.3\pm0.3$ & $100.0\pm0.7$ & $156.8\pm0.9$ & $485.1\pm1.6$ & --- & $14.0\pm0.2$ \\
		3 & $<1.1$ & $1.0\pm0.3$ & $0.6\pm0.3$ & $100.0\pm1.3$ & $159.6\pm1.3$ & $477.6\pm2.4$ & --- & $12.2\pm0.2$ \\
		4 & $1.2\pm0.4$ & $1.4\pm0.5$ & $<1.5$ & $100.0\pm1.3$ & $194.7\pm2.0$ & $573.0\pm3.4$ & $<0.7$ & $11.6\pm0.3$ \\
		5 & $<1.6$ & $4.3\pm0.6$ & $3.2\pm0.6$ & $100.0\pm1.2$ & $259.0\pm2.6$ & $765.7\pm3.5$ & $<0.8$ & $11.0\pm0.2$ \\
		6 & $<2.8$ & $<2.9$ & $<3.0$ & $100.0\pm2.3$ & $180.9\pm3.1$ & $539.9\pm4.2$ & $<1.5$ & $12.3\pm0.6$ \\
		7 & $2.3\pm0.4$ & $0.9\pm0.3$ & $<1.0$ & $100.0\pm1.1$ & $162.0\pm1.5$ & $476.6\pm2.5$ & $<0.3$ & $11.1\pm0.1$ \\
		8 & $<7.4$ & $<6.9$ & $<6.1$ & $100.0\pm2.4$ & $68.1\pm2.4$ & $196.2\pm3.4$ & $<2.1$ & $7.6\pm0.7$ \\
		9 & $<18.0$ & $<17.5$ & $<13.4$ & $100.0\pm7.5$ & $127.8\pm6.6$ & $362.7\pm8.4$ & $<8.5$ & $5.2\pm2.3$ \\
		10 & $7.3\pm2.6$ & $<10.0$ & $<7.7$ & $100.0\pm5.1$ & $70.5\pm5.1$ & $205.9\pm5.5$ & --- & $9.8\pm1.0$ \\ \hline \hline
		\# & [O\,{\sc i}]$\lambda$6300 & [S\,{\sc iii}]$\lambda$6312 & [N\,{\sc ii}]$\lambda$6548 & H$\alpha$ & [N\,{\sc ii}]$\lambda$6584 & He\,{\sc i}\,$\lambda$6678 & [S\,{\sc ii}]$\lambda$6716 & [S\,{\sc ii}]$\lambda$6731 \\
		(1) & (18) & (19) & (20) & (21) & (22) & (23) & (24) & (25) \\ \hline
		1 & $3.4\pm1.2$ & $2.3\pm1.0$ & $<2.7$ & $264.2\pm1.9$ & $2.5\pm0.5$ & $3.1\pm0.9$ & $4.3\pm0.7$ & $3.2\pm0.6$ \\
		2 & $1.6\pm0.1$ & $1.9\pm0.1$ & $1.0\pm0.1$ & $279.1\pm0.6$ & $3.3\pm0.1$ & $2.9\pm0.1$ & $6.2\pm0.2$ & $6.0\pm0.2$ \\
		3 & $2.9\pm0.1$ & $1.6\pm0.1$ & $1.6\pm0.1$ & $322.3\pm0.7$ & $3.1\pm0.1$ & $3.1\pm0.1$ & $14.2\pm0.2$ & $10.0\pm0.2$ \\
		4 & $3.9\pm0.3$ & $2.3\pm0.3$ & $2.3\pm0.2$ & $313.4\pm1.1$ & $6.2\pm0.3$ & $2.0\pm0.3$ & $14.6\pm0.4$ & $10.6\pm0.4$ \\
		5 & $1.5\pm0.3$ & $1.6\pm0.2$ & $1.5\pm0.3$ & $306.9\pm1.4$ & $1.5\pm0.3$ & $3.4\pm0.4$ & $5.9\pm0.3$ & $4.7\pm0.3$ \\
		6 & $2.1\pm0.4$ & $1.5\pm0.4$ & $<1.7$ & $322.0\pm2.0$ & $2.2\pm0.5$ & $3.3\pm0.5$ & $9.8\pm0.5$ & $7.2\pm0.5$ \\
		7 & $2.9\pm0.1$ & $1.9\pm0.1$ & $2.0\pm0.1$ & $325.8\pm0.7$ & $5.3\pm0.2$ & $1.7\pm0.1$ & $13.4\pm0.2$ & $10.3\pm0.2$ \\
		8 & $2.2\pm0.8$ & $<2.4$ & $<2.6$ & $261.2\pm1.3$ & $<2.5$ & $<3.3$ & $5.7\pm0.7$ & $5.1\pm1.2$ \\
		9 & $<6.7$ & $<7.1$ & $<8.7$ & $273.5\pm3.8$ & $<10.4$ & $<7.9$ & $<7.8$ & $5.3\pm2.6$ \\
		10 & $2.9\pm0.7$ & $<3.0$ & $<3.1$ & $302.1\pm2.1$ & $<4.2$ & $4.0\pm1.5$ & $5.1\pm1.1$ & $4.0\pm1.5$ \\ \hline \hline
		\# & He\,{\sc i}\,$\lambda$7065 & [Ar\,{\sc iii}]$\lambda$7136 & [O\,{\sc ii}]$\lambda$7320 & [O\,{\sc ii}]$\lambda$7330 & \multicolumn{2}{c}{\tcrg{$F({\rm H}\beta)$}} & & \\
		& & & & & \multicolumn{2}{c}{$10^{-17}$ erg s$^{-1}$ cm$^{-2}$} & & \\
		(1) & (26) & (27) & (28) & (29) & \multicolumn{2}{c}{(30)} & & \\ \hline
		1 & $1.5\pm0.6$ & $2.9\pm1.4$ & $<2.7$ & $<2.1$ & \multicolumn{2}{c}{$52.8\pm1.6$} & & \\
		2 & $2.4\pm0.1$ & $5.7\pm0.2$ & $1.1\pm0.1$ & $1.1\pm0.2$ & \multicolumn{2}{c}{$519.5\pm3.6$} & & \\
		3 & $2.6\pm0.2$ & $6.4\pm0.2$ & $2.6\pm0.2$ & $2.2\pm0.2$ & \multicolumn{2}{c}{$362.4\pm4.7$} & & \\
		4 & $<0.8$ & $<1.1$ & $<0.6$ & $<0.6$ & \multicolumn{2}{c}{$152.4\pm2.0$} & & \\
		5 & $4.4\pm0.4$ & $5.7\pm0.3$ & $<0.5$ & $0.9\pm0.4$ & \multicolumn{2}{c}{$209.0\pm2.5$} & & \\
		6 & $4.8\pm0.9$ & $5.4\pm0.5$ & $<1.3$ & $1.9\pm0.7$ & \multicolumn{2}{c}{$92.9\pm2.1$} & & \\
		7 & $2.8\pm0.2$ & $6.5\pm0.2$ & $2.0\pm0.2$ & $1.6\pm0.2$ & \multicolumn{2}{c}{$375.5\pm4.1$} & & \\
		8 & $1.5\pm0.6$ & $<4.7$ & $<2.8$ & $<1.8$ & \multicolumn{2}{c}{$109.4\pm2.6$} & & \\
		9 & $<13.6$ & $<12.1$ & $<6.6$ & $<6.1$ & \multicolumn{2}{c}{$12.6\pm0.9$} & & \\
		10 & $2.5\pm0.8$ & $<5.2$ & $5.8\pm1.8$ & $<4.0$ & \multicolumn{2}{c}{$22.9\pm1.2$} & & \\ \hline
    \end{tabular}
    \end{center}
    \tablecomments{(1): Number. (2): ID. (3)--(\tcrg{29}): Dust-corrected flux normalized by the H$\beta$ flux. The upper limits represent ${\rm S/N}=3$ levels. The symbols `---' indicate the lack of data because the emission lines fall into the wavelength gap between the LRIS blue and red channels. \tcrg{(30): Aperture-corrected flux of H$\beta$. Note that the errors shown in this table include only statistical errors. The LRIS EMPG candidates with EMPG-tails (\#2, 3, 4, 5, 6, 7, 9, and 10; Section \ref{subsec:lris}) have fluxes with additional systematic lower errors (10\% of the fluxes) originated from the potential contaminations from the EMPG-tails (Section \ref{subsec:flux}).}}
\end{table*}

\begin{table*}[t]
    \begin{center}
    \caption{Fundamental properties of the LRIS EMPG candidates}
    \label{tab:prop}
    \begin{tabular}{cccccccc} \hline \hline
		\# & ID & Redshift & $12+\log(\rm O/H)$ & EW$_{0}$(H$\beta$) & $E(B-V)$ & $T_{\rm e}$ & \tcrg{$n_{\rm e}$} \\
		& & & & \AA & mag & 10$^{4}$ K & cm$^{-3}$\\
		(1) & (2) & (3) & (4) & (5) & (6) & (7) & (8) \\ \hline
		1 & J0156$-$0421 & 0.04907 & $7.48^{+0.07}_{-0.06}$ & $>76.0$ & $0.00^{+0.04}_{-0.00}$ & $1.79^{+0.12}_{-0.11}$ & $127^{+539}_{-127}$ \\
		2 & J0159$-$0622 & 0.00852 & $7.68^{+0.04}_{-0.03}$ & $127.0^{+19.8+63.5}_{-19.8-12.7}$ & $0.10\pm0.01$ & $1.65^{+0.04}_{-0.04}$ & $735^{+196}_{-179}$ \\
		3 & J0210$-$0124 & 0.01172 & $7.76^{+0.03}_{-0.03}$ & $78.9^{+4.6+39.5}_{-4.6-7.9}$ & $0.06\pm0.01$ & $1.62^{+0.04}_{-0.05}$ & $4^{+63}_{-4}$ \\
		4 & J0214$-$0243 & 0.02860 & $7.96^{+0.04}_{-0.04}$ & $106.5^{+8.2+53.3}_{-8.2-10.7}$ & $0.03\pm0.01$ & $1.43^{+0.05}_{-0.04}$ & $64^{+96}_{-64}$ \\
		5 & J0226$-$0517 & 0.04386 & $7.78^{+0.04}_{-0.04}$ & $136.6^{+12.5+68.3}_{-12.5-13.7}$ & $0.14\pm0.02$ & $1.78^{+0.05}_{-0.05}$ & $268^{+223}_{-177}$ \\
		6 & J0232$-$0248 & 0.04336 & $7.73^{+0.04}_{-0.04}$ & $109.0^{+18.1+54.5}_{-18.1-10.9}$ & $0.08\pm0.02$ & $1.69^{+0.06}_{-0.06}$ & $76^{+192}_{-76}$ \\
		7 & J1608+4337 & 0.02896 & $7.75^{+0.04}_{-0.04}$ & $83.4^{+2.3+41.7}_{-2.3-8.3}$ & $0.08\pm0.01$ & $1.62^{+0.05}_{-0.05}$ & $155^{+107}_{-78}$ \\
		8 & J2236+0444 & 0.02870 & $7.05^{+0.17}_{-0.12}$ & $>44.8$ & $0.28\pm0.05$ & $2.22^{+0.30}_{-0.32}$ & $548^{+902}_{-516}$ \\
		9 & J2321+0125 & 0.05639 & $7.28^{+0.29}_{-0.16}$ & $>39.6$ & $0.05^{+0.09}_{-0.05}$ & $2.23^{+0.39}_{-0.48}$ & --- \\
		10 & J2355+0200 & 0.01231 & $7.15^{+0.20}_{-0.13}$ & $78.0^{+20.5+39.0}_{-20.5-7.8}$ & $0.09\pm0.05$ & $2.06^{+0.31}_{-0.35}$ & $214^{+1934}_{-214}$ \\ \hline
    \end{tabular}
    \end{center}
    \tablecomments{(1): Number. \tcrb{(2): ID.} (3): Redshift of H$\beta$, whose typical uncertainty is $\mathcal{O}(10^{-6})$. (4): $12+\log(\rm O/H)$ \tcra{in the gas phase}. (5): Rest-frame equivalent width of H$\beta$. The lower limits indicate fluxes of H$\beta$ divided by continua of ${\rm S/N}=3$ levels. \tcrg{The first and second terms of the errors represent statistical and systematic errors, respectively. The systematic errors of the upper and lower errors correspond to the uncertainties originated from potential contaminations of stellar continua and H$\beta$ fluxes from EMPG-tails, respectively (Section \ref{subsec:flux}).} (6): $E(B-V)$. (7): \tcrc{Electron temperature $T_{\rm e}$}. \tcrg{(8): Electron density $n_{\rm e}$. Note that LRIS EMPG \#9 does not have $n_{\rm e}$ measurement because [S\,{\sc ii}]$\lambda$6716 is not available (see Section \ref{subsec:chem}).}}
\end{table*}

\begin{table*}[t]
    \begin{center}
    \caption{Element abundance ratios of the LRIS EMPG candidates}
    \label{tab:ratio}
    \begin{tabular}{cccccc} \hline \hline
		\# & ID & $\log(\rm Ne/O)$ & $\log(\rm Ar/O)$ & $\log(\rm N/O)$ & $\log(\rm Fe/O)$ \\
		(1) & (2) & (3) & (4) & (5) & (6) \\ \hline
		1 & J0156$-$0421 & $-0.759^{+0.021}_{-0.019}$ & $-2.39^{+0.17}_{-0.25}$ & $-1.22^{+0.10}_{-0.11}$ & $<-0.45$ \\
		2 & J0159$-$0622 & $-0.878^{+0.017}_{-0.018}$ & $-2.32^{+0.02}_{-0.02}$ & $-1.47^{+0.03}_{-0.03}$ & $-1.74^{+0.17}_{-0.28\tcrg{-0.22}}$ \\
		3 & J0210$-$0124 & $-0.688^{+0.016}_{-0.017}$ & $-2.37^{+0.02}_{-0.02}$ & $-1.72^{+0.03}_{-0.03}$ & $<-1.75$ \\
		4 & J0214$-$0243 & $-0.644^{+0.016}_{-0.018}$ & $<-3.25$ & $-1.51^{+0.03}_{-0.03}$ & $<-1.61$ \\
		5 & J0226$-$0517 & $-0.664^{+0.016}_{-0.017}$ & $-2.38^{+0.03}_{-0.03}$ & $-1.51^{+0.06}_{-0.07}$ & $<-1.27$ \\
		6 & J0232$-$0248 & $-0.692^{+0.018}_{-0.017}$ & $-2.42^{+0.05}_{-0.06}$ & $-1.79^{+0.10}_{-0.12}$ & $<-1.26$ \\
		7 & J1608+4337 & $-0.653^{+0.017}_{-0.016}$ & $-2.35^{+0.02}_{-0.03}$ & $-1.47^{+0.02}_{-0.03}$ & $-1.66^{+0.15}_{-0.22\tcrg{-0.20}}$ \\
		8 & J2236+0444 & $-0.811^{+0.063}_{-0.066}$ & $<-1.98$ & $<-1.26$ & $<-0.57$ \\
		9 & J2321+0125 & $-0.810^{+0.074}_{-0.064}$ & $<-1.76$ & $<-0.68$ & $<-0.07$ \\
		10 & J2355+0200 & $-0.813^{+0.053}_{-0.051}$ & $<-2.00$ & $<-1.13$ & $<-0.52$ \\ \hline
    \end{tabular}
    \end{center}
    \tablecomments{(1): Number. \tcrb{(2): ID.} (3)--(6): Metal-to-oxygen abundance ratios \tcra{in the gas phase}. The upper limits represent $3\sigma$ confidence levels. \tcrg{The first and second terms of the Fe/O lower errors correspond to statistical and systematic errors, respectively. The systematic errors are originated from the ICF uncertainty (Section \ref{subsec:chem}).}}
\end{table*}

\subsection{Stellar Mass Estimation} \label{subsec:mass}
We estimate stellar masses with the spectral energy distribution (SED) interpretation code, {\sc beagle} \citep{Chevallard2016}. The {\sc beagle} code calculates both the stellar continuum and the nebular emission using the stellar population synthesis code \citep{Bruzual2003} and
\tcrb{the nebular emission library of \citet{Gutkin2016} that are computed with the photoionization code {\sc cloudy} \citep{Ferland2013}.}
\tcrb{We adopt the \citet{Calzetti1994} law to the models for dust attenuation.
Assuming a constant star-formation history and the \citet{Chabrier2003} IMF, we run the {\sc beagle} code with 5 free parameters of metallicity $Z$, maximum stellar age $t_{\rm max}$, stellar mass $M_{*}$, ionization parameter $U$, and $V$-band optical depth $\tau_{\rm V}$.
To obtain a parametric range of $\tau_{\rm V}$, we use the $\pm68$\% confidence intervals of $E(B-V)$ calculated in Section \ref{subsec:flux}. 
Parametric ranges of the other 4 parameters are $Z=0.006$--0.3 ${\rm Z}_{\odot}$, $\log{(t_{\rm max}/{\rm yr})}=4.0$--9.0, $\log{(M_{*}/{\rm M}_{\odot})}=4.0$--9.0, and $\log{(U)}=(-2.5)$--$(-0.5)$, which are the same as those adopted in Paper III\footnote{\tcrc{When we fix $Z$ and $U$ based on our spectroscopic results of $12+\log(\rm O/H)$ and [O\,{\sc iii}]$\lambda$5007/[O\,{\sc ii}]$\lambda\lambda$3727,3729, we check that $M_{*}$ values change at most $\sim0.3$ dex, which does not change our conclusions.}}.} 
We use the redshift values obtained in Section \ref{subsec:flux}. 
\tcra{Because the HSC $y$-band image is $\sim1$ mag shallower than the other broadband images, we do not use the HSC $y$-band data but 4 broadband ($griz$) data for the SED fitting.}
\tcrc{We use {\tt cmodel} magnitudes of $griz$ bands.
Thanks to the deblending technique \citep{Huang2018}, the {\tt cmodel} magnitude represents a total magnitude of a source even if it is overlapped with other sources.}
The $griz$-{\tt cmodel} magnitudes and stellar masses of the LRIS EMPG candidates are summarized in Table \ref{tab:mass}.
\tcrb{In the table, we show only median values of the stellar masses because errors provided by the SED fitting do not include any uncertainty arising from different assumptions. This uncertainty is $\sim0.1$ dex, which is larger than a typical error of $\sim0.05$ dex provided by the SED fitting.}

\begin{table}[t]
    \begin{center}
    \caption{Stellar masses of the LRIS EMPG candidates}
    \label{tab:mass}
    \begin{tabular}{ccccccc} \hline \hline
        \# & ID & $g$ & $r$ & $i$ & $z$ & $\log(M_{*})$ \\
        & & mag & mag & mag & mag & M$_{\odot}$ \\
        (1) & (2) & (3) & (4) & (5) & (6) & (7) \\ \hline
        1 & J0156$-$0421 & 21.9 & 22.1 & 22.9 & 22.9 & \tcrb{5.7} \\
        2 & J0159$-$0622 & 18.3 & 18.6 & 19.3 & 19.2 & \tcrb{5.4} \\
        3 & J0210$-$0124 & 19.3 & 19.7 & 20.3 & 20.3 & \tcrb{5.3} \\
        4 & J0214$-$0243 & 19.1 & 19.4 & 20.0 & 20.2 & \tcrb{6.1} \\
        5 & J0226$-$0517 & 20.5 & 21.3 & 22.0 & 22.3 & \tcrb{6.4} \\
        6 & J0232$-$0248 & 20.3 & 20.8 & 21.4 & 21.7 & \tcrb{5.9} \\
        7 & J1608$+$4337 & 19.4 & 19.8 & 20.5 & 20.6 & \tcrb{6.0} \\
        8 & J2236$+$0444 & 22.3 & 22.2 & 22.8 & 22.9 & \tcrb{5.2} \\
        9 & J2321$+$0125 & 23.1 & 23.3 & 23.9 & 24.0 & \tcrb{5.4} \\
        10 & J2355$+$0200 & 21.1 & 21.1 & 22.4 & 22.4 & \tcrb{4.8} \\ \hline
    \end{tabular}
    \end{center}
    \tablecomments{(1): Number. (2): ID. (3)--(6): HSC {\tt cmodel} magnitudes. (7): Stellar mass. }
\end{table}

\section{Results and Discussions} \label{sec:result}
\subsection{Metallicity} \label{subsec:met}
\begin{figure*}[t]
    \centering
    \includegraphics[width=18.0cm]{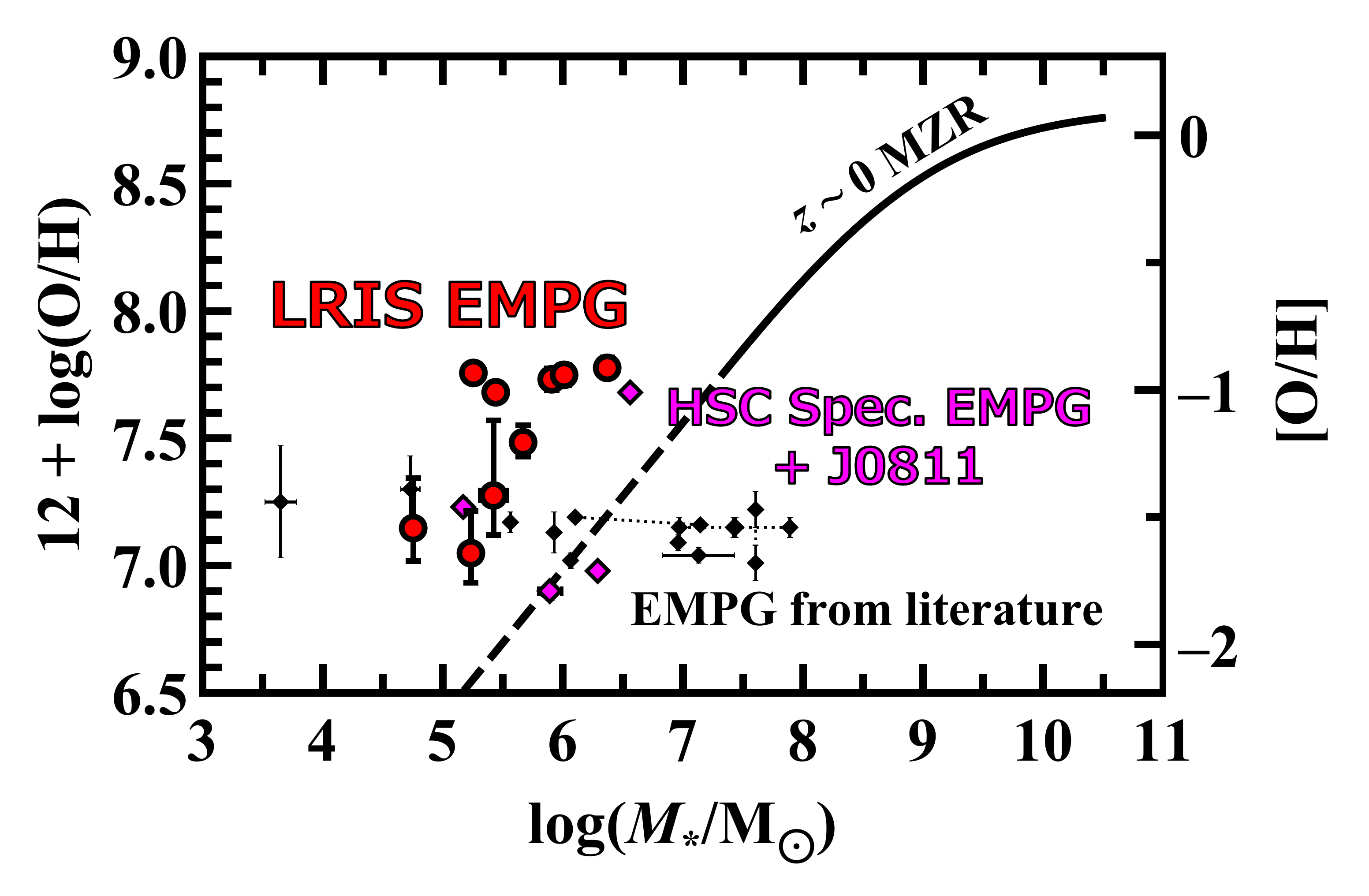}
    \caption{Mass-metallicity distribution of EMPGs. The red circles indicate the LRIS EMPGs (Section \ref{subsec:met}). 
    \tcrc{We note that a typical uncertainty of $M_{*}$ of the LRIS EMPGs is $\sim0.1$ dex (Section \ref{subsec:mass}). The magenta and black diamonds show the EMPGs listed in the upper and lower parts of Table \ref{tab:ref}, respectively.}
    The black dotted lines connecting some of the black diamonds describe that these black diamonds represent individual clumps belonging to the same EMPGs.
    The black solid curve is the $z\sim0$ mass-metallicity relation (MZR)  \tcrb{from the SDSS using the same direct-$T_{\rm e}$ method} \citep{Andrews2013}. 
    \tcrg{The black dashed line represents an extrapolation of the $z\sim0$ MZR toward lower $M_{*}$.}}
    \label{fig:mzr}
\end{figure*}

\begin{table*}[t]
    \begin{center}
    \caption{EMPGs from the literature}
    \label{tab:ref}
    \begin{tabular}{cccc} \hline \hline
        Name & $12+\log(\rm O/H)$ & $\log(M_{*}/{\rm M}_{\odot})$ & Reference \\
        (1) & (2) & (3) & (4) \\ \hline
        J1631+4426 & $6.90\pm0.03$ & $5.89^{+0.10}_{-0.09}$ & Paper I \\
        J2314+0154 & $7.23^{+0.03}_{-0.02}$ & $5.17\pm0.01$ & Paper I \\
        J2115$-$1734 & $7.68\pm0.01$ & $6.56\pm0.02$ & Paper I \\
        J0811+4730 & $6.98\pm0.02$ & $6.29\pm0.06$ & \citet{Izotov2018} \\ \hline
        AGC198691 & $7.02\pm0.03$ & 6.06 & \citet{Hirschauer2016} \\
        J1234+3901 & $7.04\pm0.03$ & $7.13\pm0.3$ & \citet{Izotov2019} \\
        J2229+2725 & $7.09\pm0.03$ & 6.96 & \citet{Izotov2021a} \\
        LittleCub & $7.13\pm0.08$ & 5.93 & \citet{Hsyu2017} \\
        LeoP & $7.17\pm0.04$ & 5.56 & \citet{Skillman2013} \\
        J1005+3722 & $7.25\pm0.22$ & $3.65\pm0.13$ & \citet{Senchyna2019} \\
        J0845+0131 & $7.30\pm0.13$ & $4.73\pm0.08$ & \citet{Senchyna2019} \\
        SBS0335\#1 & $7.01\pm0.07$ & (7.61) & \citet{Izotov2009} \\
        SBS0335\#2 & $7.22\pm0.07$ & (7.61) & \citet{Izotov2009} \\
        DDO68\#1 & ($7.15\pm0.04$) & $7.44\pm0.03$ & \citet{Sacchi2016} \\
        DDO68\#2 & ($7.15\pm0.04$) & $7.89\pm0.01$ & \citet{Sacchi2016} \\
        DDO68\#3 & ($7.15\pm0.04$) & $7.42\pm0.02$ & \citet{Sacchi2016} \\
        DDO68\#4 & ($7.15\pm0.04$) & $6.97\pm0.03$ & \citet{Sacchi2016} \\
        IZw18NW & $7.16\pm0.01$ & 7.14 & \citet{Izotov1998} \\
        IZw18SE & $7.19\pm0.02$ & 6.10 & \citet{Izotov1998} \\ \hline
    \end{tabular}
    \end{center}
    \tablecomments{(1): Name. (2): $12+\log(\rm O/H)$. (3): Stellar mass. \tcrg{We show errors of $12+\log(\rm O/H)$ and $M_{*}$ as long as they are reported in the literature.} The values in \tcra{the parentheses} indicate that the metallicities or stellar masses are not derived for each individual clumps. (4): Reference. \tcrc{In the upper part of this table, we list the HSC spectroscopic EMPGs (Paper I; Section \ref{sec:sample}) and J0811+4730 \citep{Izotov2018}. In the lower part of this table, we summarize other EMPGs with $12+\log(\rm O/H)<7.3$ (i.e., $<4$\% (O/H)$_{\odot}$).}}
\end{table*}

As listed in Table \ref{tab:ratio}, 5 out of the 10 LRIS EMPG candidates \tcrb{(\#1, 2, 8, 9, and 10)} have low metallicities of $12+\log(\rm O/H)=\tcrg{7.05}$--\tcrg{7.68}\tcra{, which} meet the EMPG criterion of $12+\log(\rm O/H)\leq7.69$. 
We also \tcrb{conclude} that 4 out of the other 5 LRIS EMPG candidate \tcrb{(\#3, 5, 6, and 7)} show a low metallicity of $12+\log(\rm O/H)=7.73$--\tcrg{7.78} (i.e., 11--12\% (O/H)$_{\odot}$). 
We thus refer the 9 ($=5+4$) LRIS EMPG candidates with $\lesssim10$\% (O/H)$_{\odot}$ as LRIS EMPGs. 
It should be noted that the other LRIS EMPG candidate \tcrb{(\#4)} still shows a low metallicity of $12+\log(\rm O/H)=7.96$ (i.e., 19\% (O/H)$_{\odot}$). 
We emphasize that \tcrb{2} out of the LRIS EMPGs \tcrb{(\#8 and 10)} show extremely-low metallicities of $12+\log(\rm O/H)=\tcrg{6.93}$--\tcrg{7.35} (i.e., \tcrg{1.7}--\tcrg{4.6}\% (O/H)$_{\odot}$) \tcrg{including the 1$\sigma$ uncertainties.
The 2 LRIS EMPGs are thus the most metal-poor galaxies ever reported.}

Figure \ref{fig:mzr} illustrates mass-metallicity distributions of the LRIS EMPGs, the HSC spectroscopic EMPGs (see Section \ref{sec:sample}), and all the other EMPGs with $12+\log(\rm O/H)<7.3$ (i.e., $<4$\% (O/H)$_{\odot}$) determined by the direct-$T_{\rm e}$ method \tcra{taken from the literature}. 
The EMPGs from the literature are \tcra{listed} in Table \ref{tab:ref}. 
\tcrc{The mass-metallicity relation (MZR) of typical SFGs at $z\sim0$ is well investigated and explained by the equilibrium of gas inflow/outflow and metal production by stars \citep[e.g.,][]{Lilly2013}.
However, we confirm that EMPGs show a wide range of $M_{*}$ from $10^{4}$ to $10^{8}$ M$_{\odot}$, which implies that the equilibrium is not maintained at the low-mass (low-metallicity) end of the mass-metallicity distribution.
Some EMPGs including many of the LRIS EMPGs lie above the extrapolation of the $z\sim0$ MZR.
Such EMPGs may be dominated by internal metal productions or outflows (Paper I), and be in the stage of the transition from gas-rich dwarf irregulars to gas-poor dwarf spheroidals \citep{Zahid2012}.
\tcre{As part of the on-going mid-high resolution spectroscopy survey with Magellan/MagE (EMPRESS-HRS; PI: M. Rauch), a following EMPRESS paper (Xu et al. in prep.) will report that EMPGs lying above the $z\sim0$ MZR have broad components of emission lines with velocity widths of $\sim200$ km s$^{-1}$, which may be attributed to the outflow.}
On the other hand, some of the EMPGs from the literature lie below the extrapolation of the $z\sim0$ MZR.
In such EMPGs, the contribution of metal-poor gas inflow may overwhelm the contribution of internal metal productions as discussed in \citet{Hughes2013}.
%We need far-UV and radio observation, which trace absorption or emission lines of the molecular gas, respectively.
}

We also find that \tcrb{LRIS EMPGs \#8 and 10} also show extremely-low stellar masses of $5\times10^{4}$--$7\times10^{5}$ M$_{\odot}$.
\tcra{LRIS EMPGs \#8 and 10 are helpful to understand the nature of galaxies in the very early formation phase because} there have been only $\sim7$ EMPGs reported so far to have both $12+\log(\rm O/H)\lesssim7.3$ and $M_{*}\lesssim10^{6}$ M$_{\odot}$. 
\tcrc{We need to explore EMPGs continuously not only to identify lowest-metallicity galaxies but also to verify whether there is a metallicity lower limit (a.k.a. metallicity floor; \citealt{Prochaska2003}) that local galaxies can take.}
\tcre{Our spectroscopic observations for the EMPRESS sample continue to address the question.
A recent follow-up identifies new EMPRESS EMPGs that have a stellar mass as low as $M_{*}\sim10^{4.7}$ M$_{\odot}$ (Nakajima et al. in prep.).
However, their metallicities do not fall below the currently-known metallicity floor of $\sim1$\% of the solar metallicity \citep[e.g.,][]{Thuan2005}, supporting a deficit of galaxies with metallicities below $\sim1$\% in the local universe.}

\subsection{Element Abundance Ratio} \label{subsec:ratio}
\begin{table*}[t]
    \begin{center}
    \caption{\tcrg{EMPGs with Fe/O measurements}}
    \label{tab:feo}
    \begin{tabular}{cccccc} \hline \hline
        Name & $12+\log(\rm O/H)$ & $\log(\rm Fe/O)$ & $\log(\rm N/O)$ & EW$_{0}$(H$\beta$) & Reference \\
        & & & & \AA & \\
		(1) & (2) & (3) & (4) & (5) & (6) \\ \hline
		LRIS EMPG \#2 & $7.68^{+0.04}_{-0.03}$ & $-1.74^{+0.17}_{-0.28-0.22}$ & $-1.47^{+0.03}_{-0.03}$ & $127^{+83}_{-33}$ & {\bf This paper} \\
		LRIS EMPG \#7 & $7.75\pm0.04$ & $-1.66^{+0.15}_{-0.22-0.20}$ & $-1.47^{+0.02}_{-0.03}$ & $83^{+44}_{-11}$ & {\bf This paper} \\ \hline
        J1631+4426 & $6.90\pm0.03$ & $-1.25^{+0.17}_{-0.31-0.22}$ & $<-1.71$ & $123.5^{+3.5}_{-2.8}$ & Papers I and II \\
        J2115$-$1734 & $7.68\pm0.01$ & $-1.64^{+0.03}_{-0.03-0.21}$ & $-1.518^{+0.009}_{-0.011}$ & $214.0^{+0.9}_{-0.8}$ & Papers I and II \\
        J0811+4730 & $6.98\pm0.02$ & $-1.06^{+0.09}_{-0.09-0.22}$ & $-1.535\pm0.044$ & $282.0\pm1.0$ & \citet{Izotov2018} \\ \hline
    \end{tabular}
    \end{center}
    \tablecomments{(1): Name. (2): $12+\log(\rm O/H)$. (3): Fe/O. (4): N/O. (5): EW$_{0}$(H$\beta$). (6): Reference.}
\end{table*}

\begin{figure*}[p]
    \centering
    \includegraphics[width=18.0cm]{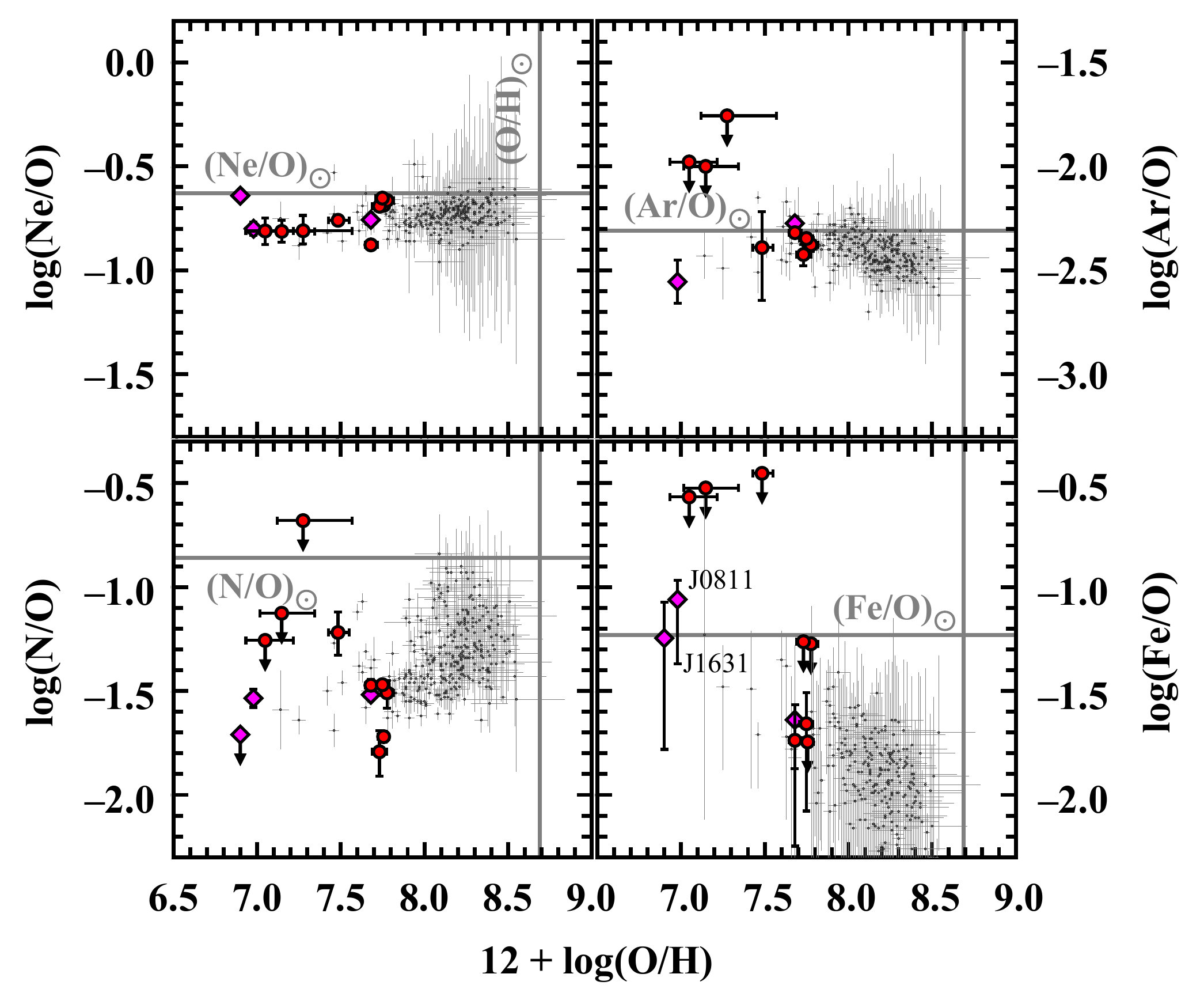}
    \caption{\tcra{Gas-phase} element abundance ratios of Ne/O (top left), Ar/O (top right), N/O (bottom left), and Fe/O (bottom right) as functions of $12+\log(\rm O/H)$. The symbols are the same as in Figure \ref{fig:mzr}. We also add local metal-poor galaxies \citep{Izotov2006} with the gray dots. \tcrg{We add 0.2-dex uncertainties to lower errors of Fe/O ratios of the local metal-poor galaxies because the Fe/O ratios are calculated from the ICF of \citeauthor{Izotov2006} (\citeyear{Izotov2006}; see Section \ref{subsec:chem}).}
    The gray solid lines represent the solar abundances \citep{Asplund2021}.}
    \label{fig:metal}
\end{figure*}

Figure \ref{fig:metal} shows element abundance ratios of Ne/O, Ar/O, N/O, and Fe/O of the LRIS EMPGs and local metal-poor galaxies of \citet{Izotov2006} (gray) as functions of metallicity. 
As shown in the top 2 panels, the LRIS EMPGs show $\alpha$-element ratios of Ne/O and Ar/O comparable to the solar abundance ratios as well as other metal-poor galaxies. 

The bottom left panel of Figure \ref{fig:metal} presents the relations between N/O and $12+\log(\rm O/H)$, illustrating that local metal-poor galaxies present a plateau at $\log(\rm N/O)\sim-1.6$ in the range of $12+\log(\rm O/H)\lesssim8.0$ and a positive slope at $12+\log(\rm O/H)\gtrsim8.0$.
\tcrg{Various studies such as} \citeauthor{Vincenzo2016} (\citeyear{Vincenzo2016}; hereafter V16) suggest that the plateau and the positive slope are attributed to the primary and the secondary nucleosyntheses in massive and low-mass stars, respectively.
We find that some of the LRIS EMPGs lie on the plateau.
\tcre{Such EMPGs are not likely to be affected by the chemical enrichment of low-mass stars but rather massive stars.}
\tcre{We also find that LRIS EMPGs \#3 and 6 lie below the plateau, showing very low N/O ratios of $\log(\rm N/O)=(\tcrg{-1.7})$--$(\tcrg{-1.8})$.}
The 2 LRIS EMPGs may produce oxygen selectively rather than nitrogen due to top-heavy IMFs efficiently producing massive CCSNe \tcrg{or high star-formation efficiencies (SFEs; defined as SFR normalized by gas mass) as discussed in \citet{Kumari2018}.}
%The 2 LRIS EMPGs may have top-heavy IMFs efficiently producing massive CCSNe that selectively eject oxygen rather than nitrogen.

The bottom right panel of Figure \ref{fig:metal} shows the relations between Fe/O and $12+\log(\rm O/H)$. 
\tcrb{For the present case,} we derive Fe/O for 2 out of the 9 LRIS EMPGs \tcrb{(\#2 and 7)} \tcrb{from} the [Fe\,{\sc iii}]$\lambda$4658 line \tcra{detection} (Section \ref{subsec:flux}).
%\tcre{We also pinpoint that LRIS EMPG \#3 has a strong upper limit of $\log(\rm Fe/O)<-1.76$.}
We find that LRIS EMPGs \#2 and 7 have $\log(\rm Fe/O)\sim\tcrg{-1.7}$ and ${\rm O/H}\sim10$\% (O/H)$_{\odot}$.
%\tcre{The Fe/O ratios are $\sim0.4$ dex lower than those of J1631+4426 (Papers I and II) and J0811+4730 \citep{Izotov2018}, while the O/H ratios are $\sim1$ dex higher than those of J1631+4426 and J0811+4730.}
\tcrg{Adding the 2 LRIS EMPGs to 3 EMPGs with Fe/O measurements (J1631+4426, J2115$-$1734, and J0811+4730) from the literature (Paper II; \citealt{Izotov2018}) represented by the magenta diamonds in the bottom right panel of Figure \ref{fig:metal}, we now obtain a sample of 5 EMPGs with Fe/O measurements.
We summarize properties of the 5 EMPGs in Table \ref{tab:feo}.
Even considering the offsets originated from the ICF difference between \citet{Izotov2006} and \citet{Rodriguez2005},} we confirm that J1631+4426 and J0811+4730 \tcrg{having the lowest O/H ratios of $\sim2$\% (O/H)$_{\odot}$} show \tcrg{(Fe/O)$_{\odot}$, which are higher than those of the other 3 EMPGs with $\sim10$\% (O/H)$_{\odot}$}.
%Although \citet{Rodriguez2005} report that Fe/O ratios derived from only [Fe\,{\sc iii}] lines are potentially $\sim0.2$ dex larger than those derived from both [Fe\,{\sc iii}] and [Fe\,{\sc iv}] lines, the differences are not large enough to change the conclusion. 

\tcrc{Based on the observations of quasar absorption systems, \citet{Becker2012} report that Fe/O ratios of the inter-galactic medium (IGM) at $z\sim2$--6 are almost constant at a value of $[\rm Fe/O]\sim-0.4$ (i.e., $\log(\rm Fe/O)\sim-1.6$). 
%If the $z\sim0$ IGM also has a similar Fe/O value, there should be something enhancing Fe/O ratios of J1631+4426 and J0811+4730.
\tcre{If the $z\sim0$ IGM also has a Fe/O value of $[\rm Fe/O]\sim-0.4$, we need some events such as SN explosions that enhance Fe/O ratios by $\gtrsim0.4$ dex to explain high Fe/O ratios of J1631+4426 and J0811+4730.}}
\tcrb{As discussed in Paper II, however, Type Ia SNe are less likely to be main contributors to enhance Fe/O ratios of \tcrg{J1631+4426 and J0811+4730} because of their low O/H and N/O ratios.}
We discuss the origin of the Fe/O enhancements quantitatively in Section \ref{sec:feo}. 

\section{Origin of F\lowercase{e}/O Enhancements} \label{sec:feo}
In this section, we revisit the origin of the high Fe/O ratio of \tcrg{J1631+4426 and J0811+4730 with $\sim2$\% (O/H)$_{\odot}$}\tcrb{, investigating possible contributors of HNe and PISNe that have not been discussed in Paper II (Section \ref{sec:intro}).
We also re-explore the contribution of metal-poor gas inflow into EMPGs because such inflows probably trigger star-forming activities, whose effect has not evaluated in Paper II.}
%\tcrb{We mainly discuss the Fe/O enhancements of J1631+4426 and J0811+4730, which are the most metal-poor galaxies with $\sim2$\% (O/H)$_{\odot}$.}

\subsection{\tcrg{Considerable SN}} \label{subsec:sn}
\tcrg{We quantitatively explore the origin of the high Fe/O abundance ratios with Fe/O evolution models including various supernova yields.
Before explaining the models, we introduce considerable SNe that would be responsible for Fe/O ratios of EMPGs.}

Very massive stars with $\sim140$--300 M$_{\odot}$ are expected to undergo thermonuclear explosions as known as PISNe \citep{Heger2002}.
Such a massive star above $\sim140$ M$_{\odot}$ requires extremely metal-poor environments to form \tcrg{due to the efficient wind mass loss \citep[e.g.,][]{Langer2007,Hirano2014}}.
Especially, cores of stars \tcrg{with $\sim200$--300 M$_{\odot}$} are mostly transformed into $^{56}$Ni during the explosion \citep{Takahashi2018}.
The $^{56}$Ni atoms consequently decay to $^{56}$Fe \citep{Nadyozhin1994}, which largely contribute to the Fe/O enrichment. 
PISNe appear $\sim2$ Myr after the star formation, which corresponds to a lifetime of stars with $\sim300$ M$_{\odot}$ \citep{Takahashi2018}.

Massive stars with $\sim8$--100 M$_{\odot}$ evolve into neutron stars or black holes (BHs), undergoing CCSNe.
Typical CCSNe are expected to produce low Fe/O gas because $\alpha$ elements, including oxygen, are selectively created in the massive stars during the $\alpha$ reaction \citep[e.g.,][]{Nomoto2006}. 
The CCSNe emerge $\sim3$ Myr after the star formation, which corresponds to a lifetime of stars with 100 M$_{\odot}$ \citep{Portinari1998}. 

Some massive stars with $\sim30$--100 M$_{\odot}$ undergo CCSNe with explosion energies of $\gtrsim10^{52}$ erg, which is $\sim1$ dex larger than that of a typical CCSN of $\sim10^{51}$ erg.
Such CCSNe with high explosion energies are referred to as HNe \citep[e.g.,][]{Iwamoto1998}.
\tcre{The light curve model for the observed HN SN 1998bw associated
with GRB 980425 shows that the mass of Fe (mostly a decay product of
radioactive $^{56}$Ni) is $\sim0.4$ M$_{\odot}$ \citep{Nakamura2001}.
This HN model with \tcrf{$\sim3\times10^{52}$ ergs} for SN 1998bw yields
$[\rm Fe/O]\sim-0.5$.
By taking into account these observation and model of SN 1998bw, \citet{Nomoto2006,Nomoto2013} have provided yield tables giving $[\rm Fe/O]\sim-0.5$ from HNe.
However, the amount of Fe in HN models depends on the explosion energy, the progenitor mass, and the mass cut that divides the ejecta and the compact remnant.}
\citet{Umeda2008} predict that HNe with high explosion energies \tcrf{tend to produce higher Fe/O gas even} above the solar abundance \tcrc{because their high temperatures promote the nucleosynthesis of $^{56}$Ni \tcrd{that is decaying into} $^{56}$Fe.
\tcre{In addition, when we set a low value of the mass cut, HNe with normal explosion energies of \tcrf{$\sim1$--$3\times10^{52}$ erg s$^{-1}$} can also eject high Fe/O gas \tcrf{above} the solar abundance \citep{Umeda2008}.} 
\tcrf{Because HNe that eject large amount of (radioactive) $^{56}$Ni should be bright,} we refer to the HNe with high explosion energies \tcre{and/or low mass cuts} as \tcrf{bright HNe (BrHNe)}, hereafter.}
\citet{Shivvers2017} report that $\sim1$\% of observed CCSNe are HNe.
However, \citet{Maryam2020} report that galaxies hosting HNe tend to be metal-poor, which implies that HNe are preferentially born in metal-poor environments.

Low- and intermediate-mass stars evolve into white dwarfs.
If a white dwarf belongs to a binary system, the system \tcrc{may host} a Type Ia SN.
A delay time of Type Ia SNe, $\tau_{\rm Ia}$, is defined to be a time from the beginning of the star formation to an appearance of the first Type Ia SN.
We assume that the minimum possible $\tau_{\rm Ia}$ is 50 Myr \citep[e.g., ][]{Mannucci2005,Sullivan2006}, which is linked to the maximum \tcrb{zero-age main-sequence mass of stars that evolve into white dwarfs ($\sim8$ M$_{\odot}$)}\footnote{Using observational data of SNe hosted by galaxies with old stellar populations, \citet{Totani2008} report a reasonable $\tau_{\rm Ia}$ range of 0.1--10 Gyr.}.
Type Ia SNe can eject high Fe/O gas above the solar abundance because carbon deflagrations in white dwarfs synthesize $^{56}$Fe.

\subsection{\tcrb{Fe/O evolution model}} \label{subsec:sm18}
\begin{table*}[t]
    \begin{center}
    \caption{\tcra{Fe/O evolution models}}
    \label{tab:sm18}
    \begin{tabular}{cccccc} \hline \hline
        Model name & Mass range & \multicolumn{3}{c}{Progenitor star with} & Yield of \\
        & M$_{\odot}$ & 9--30 M$_{\odot}$ & 30--100 M$_{\odot}$ & 140--300 M$_{\odot}$ & (Br)HN or PISN \\
        (1) & (2) & (3) & (4) & (5) & (6) \\ \hline
        \tcrd{No HN/PISN} & 9--100 & CCSN & CCSN & --- & --- \\
        HN 100\% & 9--100 & CCSN & HN & --- & \citet{Nomoto2013} \\
        \tcrc{BrHN 20\%} & 9--100 & CCSN &
        \begin{tabular}{l}
            \footnotesize{BrHN: 20\%} \\ \footnotesize{CCSN: 80\%}
        \end{tabular}
        & --- & \citet{Umeda2008} \\
        PISN & 9--300 & CCSN & CCSN & PISN & \citet{Takahashi2018} \\ \hline
    \end{tabular}
    \end{center}
    \tablecomments{(1): Model name. (2): Mass range of progenitor stars. (3)--(5): SN evolved from stars with 9--30, 30--100, and 140--300 M$_{\odot}$. We assume that stars with 100--140 M$_{\odot}$ undergo direct collapses (i.e., ejecting no gas). \tcrc{(6): Reference of the yields of HNe, BrHNe, or PISNe. Regarding CCSNe, we use the yield of \citet{Nomoto2006}.} We also adopt the Salpeter IMF and the instantaneous star-formation history for all the models.}
\end{table*}

\begin{figure*}[t]
    \centering
    \includegraphics[width=18.0cm]{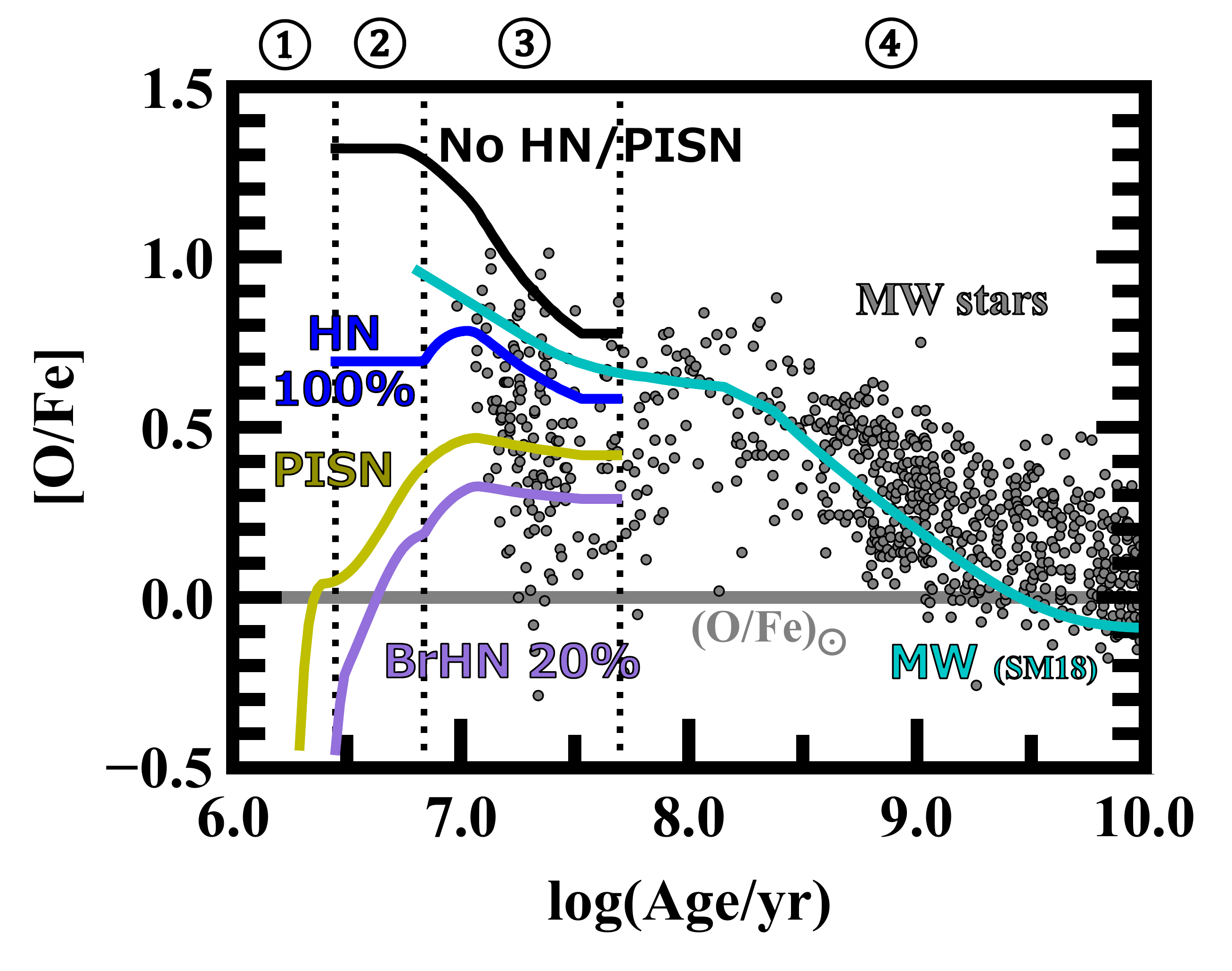}
    \caption{\tcrc{O/Fe} evolution models. 
    %The blue dotted, solid, and dashed curves indicate the HN 1, 10, and 100\% models, respectively. The PISN and the ${\rm PISN}+{\rm HN}$ 10\% models are represented by the yellow and purple curves, respectively. 
    The cyan, \tcrd{black,} blue, purple, and yellow curves indicate the MW, \tcrd{the No HN/PISN,} the HN 100\%, \tcrc{the BrHN 20\%}, and the PISN models, respectively.
    In Table \ref{tab:sm18}, we summarize the assumptions of the models mentioned above. 
    \tcrd{The gray dots show the MW stars \citep{Edvardsson1993,Reddy2003,Gratton2003,Cayrel2004,Bensby2014,Roederer2014}.}
    The gray horizontal line indicates the solar abundance of O/Fe. From the left to the right, the vertical dotted lines show lifetimes of stars with 100, 30, and 8 M$_{\odot}$, respectively.
    \tcrf{The regions \textcircled{\scriptsize 1}, \textcircled{\scriptsize 2}, \textcircled{\scriptsize 3}, and \textcircled{\scriptsize 4} correspond to eras in which only PISNe, CCSNe/(Br)HNe, CCSNe, and Type Ia SNe occur under the assumption of the instantaneous star-formation history, respectively.}
    }
    \label{fig:model}
\end{figure*}

\tcrb{To evaluate the contribution of Type Ia SNe, we use an Fe/O evolution model of \citeauthor{Suzuki2018a} (\citeyear{Suzuki2018a}; hereafter SM18).
We refer to this model as Milky Way (MW) model because the model is calibrated by observations of absorptions of the MW stars  \citep[e.g.,][]{Bensby2014}.
In the MW model, an Fe/O ratio at each age is defined as a ratio of total numbers of iron and oxygen atoms produced by all SNe that have already exploded before the age.
First, SM18 create stars based on an initial mass function (IMF) of \citet{Kroupa2001}, which have mass slopes of $-2.3$, $-1.3$, and $-0.3$ for stars with $M\geq0.5$, $0.08<M<0.5$, and $M\leq0.08$ M$_{\odot}$, respectively.
SM18 derive lifetimes of the stars as a function of star masses from \citet{Padovani1993}.
SM18 assume that all stars below 8 M$_{\odot}$ and with 9--100 M$_{\odot}$ evolve into white dwarfs and CCSNe, respecitively, after finishing their lifetimes.
The white dwarfs become Type Ia SNe following a delay-time distribution $D(t_{\rm d})$\tcrf{, which is proportional to $t_{\rm d}^{-1}$ and normalized so that the MW model reproduces ${\rm Fe/O}=(\rm Fe/O)_{\odot}$ at the age of the formation of the Sun}.
SM18 use ejecta masses (1.38 M$_{\odot}$) and element yields of Type Ia SNe predicted by \citet{Nomoto1984}.
SM18 also adopt a metallicity-dependent CCSN yield of \citet{Limongi2006} and the best-fit star-formation history of the MW\tcrf{, whose SFR continuously increases until the age of $\sim800$ Myr, and decreases to the current SFR of the MW at the age of 13.8 Gyr}.}

Because the MW model does not take into account ejecta from HNe or PISNe, we construct new Fe/O evolution models incorporating HNe or PISNe.
\tcre{Again, it is important to evaluate contributions of HNe and PISNe to the Fe/O enhancements of young metal-poor galaxies because both occur much earlier than Type Ia SNe, and also because they are preferentially born in the metal-poor environment (\citealt{Maryam2020,Langer2007}).}
We develop models whose assumptions are summarized in Table \ref{tab:sm18}.
\tcrg{The percentages appeared in the model names represent how much fraction of stars with 30--100 M$_{\odot}$ is assumed to undergo HNe (or BrHNe).}
To examine whether HNe or PISNe can contribute to the Fe/O enhancements of EMPGs, we calculate Fe/O until the age of 50 Myr (i.e., before the first Type Ia SN appears).
Other than this point, we construct the models in the same way as SM18.
We use an IMF with a mass slope of $-2.3$ (a.k.a. Salpeter IMF; \citealt{Salpeter1955})\footnote{\tcrc{The Salpeter IMF is equal to that of \citet{Kroupa2001} in the mass range of $\geq0.5$ M$_{\odot}$.}}, which is common in local star-forming galaxies.
Massive stars finish their lifetimes as main-sequence stars earlier than less-massive stars.
We derive lifetimes of the stars as a function of star masses from the combination of \citeauthor{Portinari1998} (\citeyear{Portinari1998}; for 6--120 M$_{\odot}$ stars) and \citeauthor{Takahashi2018} (\citeyear{Takahashi2018}; for 100--300 M$_{\odot}$ stars).
We calculate element ejections from CCSNe, HNe, and PISNe using the most metal-poor SN yields of \citet{Nomoto2006}, \citet{Nomoto2013}, and \citet{Takahashi2018}, respectively.
\tcrd{Because \citet{Nomoto2006} and \citet{Nomoto2013} calculate yields of CCSNe and HNe with progenitor masses of 30 and 40 M$_{\odot}$, we extrapolate the yields to CCSNe/HNe with progenitor masses of 100 M$_{\odot}$}
\tcre{To create BrHNe ejecting gas with the highest Fe/O, we use yields of \citet{Umeda2008} with the highest explosion energy and the lowest mass cut just above the Fe core at a given progenitor star mass.}
For all the models, we assume that 1) all stars with 9--30 M$_{\odot}$ explode as CCSNe, 2) all stars with 100--140 M$_{\odot}$ undergo direct collapses that eject no element, and 3) the star formation occurs at once at the beginning of the galaxy formation (i.e., instantaneous star-formation history).

\tcrb{Figure \ref{fig:model} illustrates the Fe/O evolution models\tcrd{, while we convert Fe/O to O/Fe normalized by the solar abundance ([O/Fe])}.
We find that the BrHN 20\% (purple) and PISN (yellow) models show low O/Fe (high Fe/O) ratios comparable to the solar abundance during the \textcircled{\scriptsize 1} and \textcircled{\scriptsize 2} eras.
The HN 100\% (blue), BrHN 20\%, and PISN models show maximum O/Fe (minimum Fe/O) values at $\sim$10 Myr mainly because the high-Fe/O gas produced by HNe or PISNe is diluted with the low-Fe/O gas efficiently ejected from CCSNe whose progenitor masses are $\gtrsim20$ M$_{\odot}$. 
The MW model (cyan) predicts that the O/Fe (Fe/O) value continuously decrease (increase) with the age. 
The decrease in O/Fe (increase in Fe/O) during the \textcircled{\scriptsize 4} era is mainly attributed to Type Ia SNe\footnote{\tcrc{All the models predict that O/Fe values decrease (Fe/O values increase) with the age during the \textcircled{\tiny 3} era because CCSNe whose progenitor masses are less than $\sim18$ M$_{\odot}$ (a.k.a. Type IIP SNe) produce Fe/O higher than those ejected from more-massive CCSNe (SM18).}}.}

%The gray dots in Figure \ref{fig:model} show the distribution of [O/Fe] and [Fe/H] of the MW stars \citep{Edvardsson1993,Reddy2003,Gratton2003,Cayrel2004,Bensby2014,Roederer2014}, while we convert [Fe/H] to the age using the [Fe/H] evolution models of SM18\footnote{\tcrg{To derive Fe/H of the (Br)HN and PISN models instead, we need models of the gas mass evolution of early galaxies.}}.
The gray dots in Figure \ref{fig:model} show the distribution of the MW stars \citep{Edvardsson1993,Reddy2003,Gratton2003,Cayrel2004,Bensby2014,Roederer2014}.
\tcrg{Because we cannot determine [Fe/H] evolutions of the (Br)HN/PISN models due to uncertainties of evolution models of gas masses (and thus hydrogen abundances) of EMPGs,} we convert [Fe/H] of the MW stars to the age using the [Fe/H] evolution models of SM18 instead.
The MW model reproduces the distribution of the MW stars especially during the \textcircled{\scriptsize 4} era (i.e., $[\rm Fe/H]\gtrsim-2$), which indicates that the contribution of Type Ia SNe is properly incorporated into the MW model.
In the \textcircled{\scriptsize 3} era (i.e., $[\rm Fe/H]\lesssim-2$), however, the MW model cannot explain the distribution of the MW stars with low O/Fe (high Fe/O) ratios.
The BrHN and the PISN models reproduce the distribution of the MW stars with low O/Fe ratios of $[\rm O/Fe]=0.3$--0.4, which implies that some metal-poor stars contain elements from HNe or PISNe as discussed in \citet{Aoki2014}.
We may need HNe or PISNe other than Type Ia SNe or CCSNe to reproduce the chemical enrichment of galaxies in the early formation phase.

Finally, we convert the \tcrf{model} age to EW$_{0}$(H$\beta$) because the stellar age used in the models is not an observable. 
Using the {\sc beagle} code calculating both the stellar continuum and the nebular emission (cf. Section \ref{subsec:mass}), Paper I derives EW$_{0}$(H$\alpha$) as a function of stellar age and $12+\log(\rm O/H)$ under the assumption of the constant star formation.
\tcrc{Basically, EW$_{0}$(H$\alpha$) increases as $12+\log(\rm O/H)$ decreases because \tcrf{metal-poor} stars make stellar continua harder \citep[e.g.,][]{Tumlinson2000}.}
To remove the $12+\log(\rm O/H)$ dependency, we assume a relation between O/H and stellar age of the Milky Way \tcrb{model} (SM18). Using an empirical relation of ${\rm EW}_{0}({\rm H}\alpha)=5.47\times{\rm EW}_{0}({\rm H}\beta)$ (Paper I), we obtain a relation between EW$_{0}$(H$\beta$) and the stellar age.

\tcrc{Here we evaluate the uncertainty in the conversion from the stellar age to EW$_{0}$(H$\beta$).
Paper I and \citet{Izotov2018} report that J1631+4426 and J0811+4730 have stellar ages of 50 and 3.3 Myr and EW$_{0}$(H$\beta$) of 120 and 280 \AA, respectively.
Using the relation between EW$_{0}$(H$\beta$) and the stellar age, we find that the stellar ages of 50 and 3.3 Myr correspond to EW$_{0}$(H$\beta$) of 90 and 380 \AA, respectively.
The stellar ages inferred from EW$_{0}$(H$\beta$) are different from those from the literature by $\sim0.13$ dex.
Thus, we add the $\pm0.13$-dex uncertainties to EW$_{0}$(H$\beta$) values of the Fe/O evolution models.}

\tcrc{Although the models provide total ($={\rm gas}+{\rm dust}+{\rm stellar}$) Fe/O ratios, we note that the gas-phase Fe/O ratios of dust-poor and low-$M_{*}$ EMPGs are expected to be comparable to the total Fe/O ratios (and thus the Fe/O ratios predicted by the models) because we can ignore the amount of ejecta of SNe trapped in dust grains and stars.}

\subsection{\tcrb{Possible scenario}} \label{subsec:scenario}
To explain the Fe/O enhancements, we investigate \tcrb{3 scenarios of 1) Type Ia SN, 2) gas dilution and episodic star formation, and 3) HN, BrHN, or PISN}. 

\subsubsection{\tcrb{Type Ia SN}} \label{subsubsec:IaSN}
\begin{figure*}[t]
    \centering
    \includegraphics[width=18.0cm]{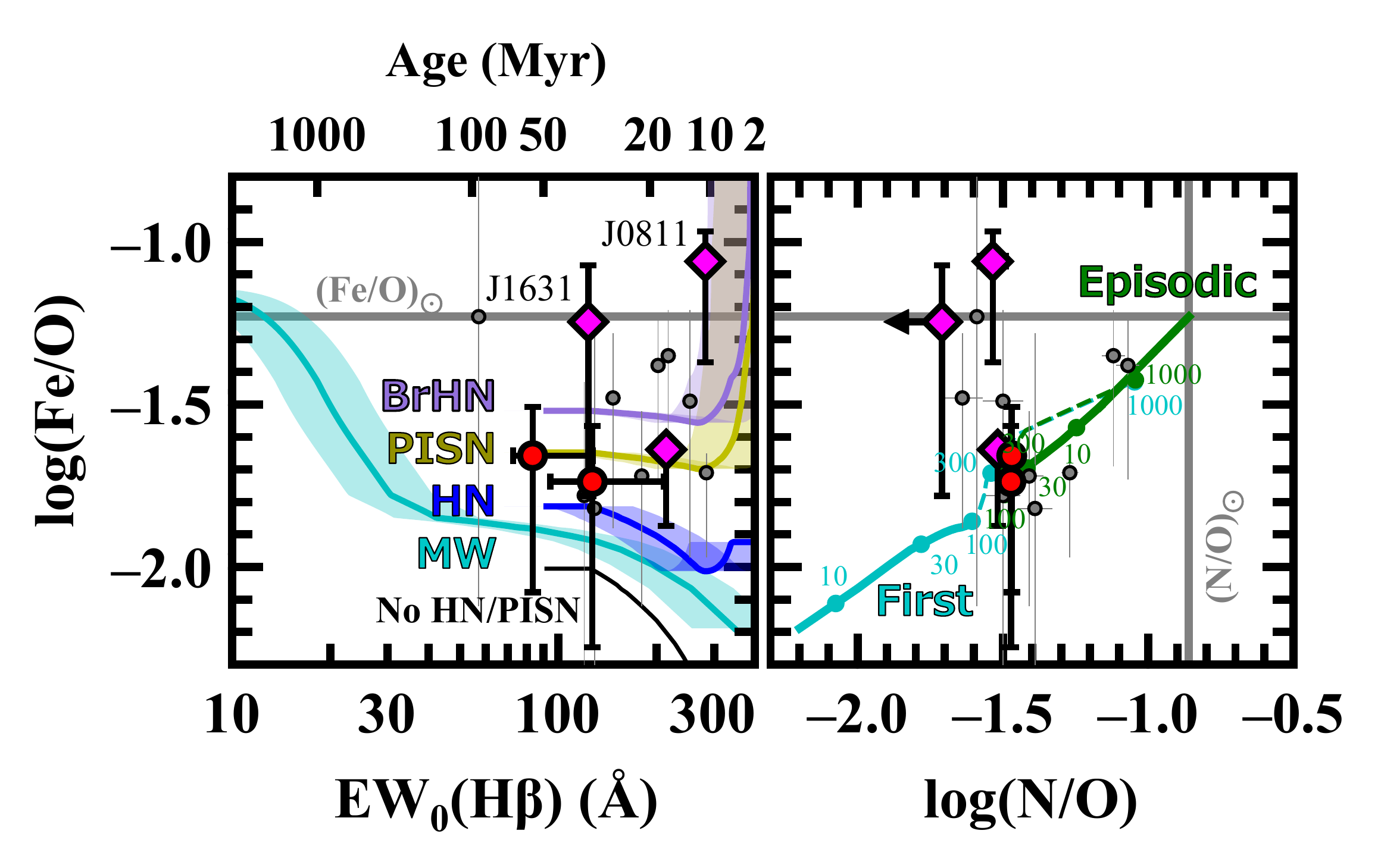}
    \caption{Fe/O ratio as a function of ${\rm EW}_{0}({\rm H}\beta)$ and N/O described in the left and right panels, respectively. The gray circles indicate local metal-poor galaxies \citep{Izotov2006}\tcrb{, which are the same as in Figure \ref{fig:metal} but limited to those} with $12+\log(\rm O/H)\leq7.69$. 
    \tcrc{Regarding the local metal-poor galaxies \citep{Izotov2006}, we only show errors of Fe/O.}
    The other symbols are the same as in Figure \ref{fig:metal}, but \tcre{limited to those with Fe/O measurements or strong upper limits of Fe/O}. 
    In the left panel, the cyan, blue, purple, and yellow curves with the shaded regions show the MW, the HN 100\%, \tcrc{the BrHN 20\%}, and the PISN models, respectively.
    \tcrd{The black curve indicates the No HN/PISN model.}
    In the right panel, the cyan and green curves represent evolution tracks \tcrb{of Fe/O and N/O} when first and episodic starbursts occur in a galaxy, respectively. The numbers accompanied by the curves indicate the ages in the unit of Myr. \tcrb{We draw the evolution tracks with solid and dashed lines before and after 100 Myr, respectively, because all the EMPGs shown in the left panel have ages $\lesssim100$ Myr. To predict Fe/O and N/O ratios, we use the MW model (SM18) and the V16 model, respectively.} The gray lines show the solar abundances \citep{Asplund2021}.}
    \label{fig:fe_ew}
\end{figure*}
\tcrb{First, we quantitatively check whether Type Ia SNe are not responsible for the Fe/O enhancements as concluded by Paper II.}
\tcrb{Below, we use the MW model (SM18)\tcre{, in which} only Type Ia SNe can eject iron-rich gas.
In the left panel of Figure \ref{fig:fe_ew}, we show the MW model represented by the cyan curve with the shaded region, respectively.}
The Fe/O value increases as the EW$_{0}$(H$\beta$) value decreases (and thus the age increases) especially in the range of EW$_{0}$(H$\beta)\lesssim50$ \AA\ (i.e., $\gtrsim100$ Myr) because of the appearance of Type Ia SNe. 

%We find that most of the EMPGs show Fe/O ratios higher than that predicted by the MW model at a given EW.
\tcrb{We find that J0811+4730 shows the Fe/O value significantly higher than the MW model prediction with a little uncertainty.
J0811+4730 has a large EW$_{0}$(H$\beta$) of $\sim300$ \AA\ corresponding to an extremely-young age of $\sim10$ Myr, which suggests that Type Ia SNe, whose minimum possible delay time is 50 Myr (Section \ref{subsec:sn}), cannot contribute to the Fe/O enhancement of J0811+4730.}
Thus, we can rule out the first scenario for J0811+4730.
We can also reject the possibility that the Fe/O value of J1631+4426 is consistent with the MW model with \tcrg{at least $\sim1\sigma$} level.
\tcrb{In addition, the Fe/O excess of J1631+4426 from the MW model may become larger because J1631+4426 has an EMPG-tail (Paper III\tcrf{; see Section \ref{subsec:lris}}), which potentially makes the EW of J1631+4426 underestimated \tcrg{by a factor of 2} (Section \ref{subsec:flux}).
We thus conclude that the Fe/O enhancement of J1631+4426 is not likely to be attributed to Type Ia SNe.}

\tcrb{Relations between Fe/O and N/O of \tcrg{J1631+4426 and J0811+4730} also support the conclusion mentioned above.
To estimate the N/O enrichment, we utilize a chemical evolution model of V16.
\tcrc{Because V16 implement the primary nucleosynthesis of nitrogen in massive stars \citep{Romano2010},} the V16 model can reproduce the observed relation between N/O and $12+\log(\rm O/H)$ of local star-forming galaxies \tcrc{within a wide metallicity range of $12+\log(\rm O/H)\sim7.2$--8.7.}
In the right panel of Figure \ref{fig:fe_ew}, we show evolution tracks of the relation between Fe/O and N/O based on \tcra{the MW model (SM18) and the V16 model represented by the cyan solid curve.}
We find that J0811+4730 and J1631+4426 have very high Fe/O and low N/O ratios, which cannot be reproduced by the first star-formation model.
We conclude again that Type Ia SNe are not likely to enhance the Fe/O ratios of J0811+4730 and J1631+4426.}

\tcrg{We note that the EMPGs other than J0811+4730 or J1631+4426 also have young ages of $<100$ Myr inferred from their large EW$_{0}$(H$\beta$) as shown in the left panel of Figure \ref{fig:fe_ew}.
However, none of the EMPGs are consistent with the first star-formation model with ages less than 100 Myr as illustrated in the right panel of Figure \ref{fig:fe_ew}.
This indicates that Type Ia SNe are not responsible for Fe/O ratios of typical EMPGs.}

\subsubsection{\tcrb{Gas dilution and episodic star formation}} \label{subsubsec:epi}
The second scenario is a combination of gas dilution and episodic star formation caused by primordial gas inflow. 
Paper III reports that more than 80\% of EMPGs have EMPG-tails.
\citet{SanchezAlmeida2015} report that EMPG-tails have O/H ratios $\sim1$ dex larger than those of EMPGs, which suggests that metal-poor gas accretions onto EMPG-tails trigger star-formation activities of EMPGs.
If EMPGs and EMPG-tails originally share the same gas, it is possible that the EMPGs have high Fe/O gas produced by past star formations in the EMPG-tails.

In this section, we assume that EMPGs originally have all element abundances \tcrc{equal} to the solar abundances based on the fact that a typical EMPG-tail has ${\rm O/H}\sim(\rm O/H)_{\odot}$ \citep{SanchezAlmeida2015}.
Then, we investigate how element abundance ratios of the EMPGs change after the metal-poor gas inflow into the EMPGs.
\tcrg{If the $z\sim0$ IGM also has an Fe/O ratio of $[\rm Fe/O]\sim-0.4$ comparable to those at $z\gtrsim2$ \citep{Becker2012}, an IGM-gas inflow onto an EMPG may decrease Fe/O of the EMPG. However, the decline is probably negligible because the IGM metallicity in the local universe may have a lower limit of $\sim10^{-2}$ Z$_{\odot}$ \citep[e.g.,][]{Thuan2005}.}
For the sake of simplicity, we assume primordial (i.e., only hydrogen) gas inflow.
Just after the original gas components of the EMPGs are diluted with the primordial gas, metal-to-hydrogen ratios such as $12+\log(\rm O/H)$ decreases while Fe/O ratios remain comparable to (Fe/O)$_{\odot}$. 
However, N/O ratios should also be similar to (N/O)$_{\odot}$, whereas the EMPGs have N/O ratios significantly lower than (N/O)$_{\odot}$ as described in the \tcra{left bottom} panel of Figure \ref{fig:metal}. 
Thus, the dilution of the solar-metallicity gas with the primordial gas cannot explain the Fe/O enhancements of the EMPGs \tcra{by itself} \tcrb{as discussed in Paper II}. 

\tcra{However, the primordial gas inflow can trigger the episodic star formation, which potentially impacts element abundance ratios of the whole EMPG due to its low stellar mass. 
Especially, the N/O ratio is expected to decrease until $\sim100$ Myr after the episodic star formation because massive stars produce low N/O gas of $\sim20$\% (N/O)$_{\odot}$ (V16).
Thus}, we investigate the contribution of the episodic star formation\tcrb{, which has not been investigated in Paper II}. 
\tcrc{As in Section \ref{subsubsec:IaSN}, we adopt the MW model and the V16 model to predict Fe/O and N/O evolutions, respectively.}
Assuming that the inflow \tcrb{quintuples the} gas mass of the galaxy after finishing the first major star formation, we calculate the evolution track of the episodic starburst as shown in the right panel of Figure \ref{fig:fe_ew} with the green solid curve. 

\tcrb{Regarding J1631+4426 and J0811+4730, we identify that even the episodic star-formation model cannot reproduce the high Fe/O, the low N/O, and the young ages of $\lesssim50$ Myr at the same time.
Of course, we can create episodic star-formation models that satisfy the high Fe/O, the low N/O, and the young ages of the EMPGs by arbitrarily assuming that the EMPGs originally have high Fe/O and low N/O ratios.
However, such an assumption is unlikely to be plausible because the first burst model at any age does not reproduce high Fe/O and low N/O ratios simultaneously.}
We conclude that the second scenario can be ruled out \tcrg{for J1631+4426 and J0811+4730} even if the inflow triggers the episodic starburst. 

\tcrg{We note that many of the EMPGs other than J1631+4426 or J0811+4730 have relatively high Fe/O and N/O ratios comparable to those predicted by the episodic star-formation model within the range of $<100$ Myr.
These EMPGs may have old gas populations that are already affected by Type Ia SNe and AGB stars as discussed in \citet{Almeida2016}.}

\subsubsection{HN, \tcrc{BrHN}, or PISN} \label{subsubsec:pisn}
The third scenario is the contribution of HNe, BrHNe, or PISNe\tcra{, which has not been investigated in Paper II}. 
The left panel of Figure \ref{fig:fe_ew} illustrates the models containing HNe, BrHNe, or PISNe (Table \ref{tab:sm18}).
We find that either the \tcrc{BrHN} 20\% (purple) or the PISN (yellow) models can reproduce the relations between Fe/O and EW$_{0}({\rm H}\beta)$ of J1631+4426 and J0811+4730.
\tcrc{We also find that we cannot explain the Fe/O enhancements even with the HN 100\% model (blue), which implies that HNe with relatively-low explosion energies of $\sim10^{52}$ erg \tcre{and reasonable mass cuts} are not responsible for the Fe/O enhancements.}
We conclude that \tcrc{BrHNe} or PISNe can contribute to the Fe/O enhancements of J1631+4426 and J0811+4730.
\tcre{The N/O ratio potentially isolate BrHNe from PISNe.
We need BrHN and PISN yields that plausibly calculate the primary nucleosynthesis of nitrogen as implemented in V16.}

\tcrg{We note that most of the EMPGs in Figure \ref{fig:fe_ew} are also in agreement with either the BrHN 20\% or the PISN models.
This may trace the (past) presence of BrHNe or PISNe in the EMPGs (with moderate Fe/O ratios). Some of the EMPGs have low Fe/O ratios comparable to that predicted by the HN 100\% model. Such EMPGs may be affected by (normal) HNe.}

\tcrb{One may wonder whether the presence of BrHNe or PISNe conflicts with the positive trend between O/Fe and Fe/H ratios of the MW stars \tcrc{(as well as stars in satellite galaxies of the MW; e.g., \citealt{Pompeia2008})}, which is an observational support of a cosmic clock.
However, the positive trend is clear only within the range of $[\rm Fe/H]\gtrsim-2$, which corresponds to the formation age of the MW of $\gtrsim50$ Myr in Figure \ref{fig:model}.
\tcre{Stars below $[\rm Fe/H]\lesssim-2$ (the formation age of the MW less than $\sim50$ Myr) show a wide range of [O/Fe] from $\sim-0.3$ to 1.0, which allows the presence of BrHNe or PISNe (see Section \ref{subsec:sm18}).}
}

\subsubsection{\tcrb{Conclusion}} \label{subsubsec:con}
We have \tcrb{investigated} the 3 scenarios that can explain the Fe/O enhancements of \tcrg{J1631+4426 and J0811+4730 with $\sim2$\% (O/H)$_{\odot}$ and low N/O ratios}. 
%Especially for J0811+4730 and J1631+4426, the third scenario of BrHNe or PISNe is most likely among the 3 scenarios.
%We conclude again that \tcrc{BrHNe} or PISNe potentially contribute to the Fe/O enhancements of EMPGs.
We conclude that the Fe/O enhancements are not likely to be explained by the Type Ia SN or the episodic star-formation scenarios but an inclusion of BrHNe and/or PISNe.
This conclusion implies that first galaxies at $z\sim10$ with metallicities of 0.1--1\% Z$_{\odot}$ \citep{Wise2012a} could also have high Fe/O ratios because HNe and PISNe are preferentially produced in metal-poor environments (Section \ref{subsec:sn}).
Our conclusion also suggests that galaxies with high Fe/O ratios are not necessarily old enough to be affected by Type Ia SNe.
This infers that Fe/O would not serve as a cosmic clock in primordial galaxies \tcre{because not Type Ia SNe but HNe or PISNe are likely to be responsible for the Fe/O enhancements in young metal-poor galaxies}.

\tcrg{We note that Fe/O ratios of the EMPGs other than J1631+4426 or J0811+4730 can be generally explained by either episodic star formation or (Br)HNe/PISNe due to their moderate Fe/O and N/O ratios compared with J1631+4426 and J0811+4730.
Primordial galaxies such as J1631+4426 and J0811+4730 are thus important to verify the presence of BrHNe or PISNe.}

\section{Summary} \label{sec:sum}
We present element abundance ratios of local EMPGs, which are expected to be galaxies in the early formation phase. 
We conduct spectroscopic follow-up observations for 13 faint EMPG candidates selected by EMPRESS with Keck/LRIS.
We newly identify 9 EMPGs with ${\rm O/H}=2.2$--12\% (O/H)$_{\odot}$ at $z=0.009$--0.057. 
Notably, \tcrb{2} out of the 9 EMPGs have extremely-low stellar masses and oxygen abundances of $5\times10^{4}$--$7\times10^{5}$ M$_{\odot}$ and 2--3\% (O/H)$_\odot$, respectively, indicative that the \tcrb{2} EMPGs are galaxies in the very early formation phase.
Comparing nucleosynthesis models with representative EMPGs, we pinpoint \tcrg{J1631+4426 and J0811+4730 having the lowest O/H ratios of $\sim2$\% (O/H)$_{\odot}$}, whose high Fe/O and low N/O ratios cannot be explained by Type Ia supernovae (SNe) or episodic star formation but bright hypernovae (BrHNe; Section \ref{subsec:sn}) and/or pair-instability SNe (PISNe).
\tcrb{Because HNe and PISNe are preferentially produced in metal-poor environments, primordial galaxies at $z\sim10$ potentially have high Fe/O values as well as the EMPGs.
We also suggest that the Fe/O ratio may not serve as a cosmic clock for primordial galaxies.}

\acknowledgments

  This paper includes data gathered with the 10-meter Keck Telescope located at W. M. Keck Observatory, Hawaii. We thank the staff of Keck Observatory for their help with the observations. 
  The Hyper Suprime-Cam (HSC) collaboration includes the astronomical communities of Japan and Taiwan, and Princeton University. The HSC instrumentation and software were developed by the National Astronomical Observatory of Japan (NAOJ), the Kavli Institute for the Physics and Mathematics of the Universe (Kavli IPMU), the University of Tokyo, the High Energy Accelerator Research Organization (KEK), the Academia Sinica Institute for Astronomy and Astrophysics in Taiwan (ASIAA), and Princeton University. Based on data collected at the Subaru Telescope and retrieved from the HSC data archive system, which is operated by Subaru Telescope and Astronomy Data Center at NAOJ. This work was supported by the joint research program of the Institute for Cosmic Ray Research (ICRR), University of Tokyo. 
  The Cosmic Dawn Center is funded by the Danish National Research Foundation under grant No. 140. 
  S.F. acknowledges support from the European Research Council (ERC) Consolidator Grant funding scheme (project ConTExt, grant No. 648179). 
  This project has received funding from the European Union's Horizon 2020 research and innovation program under the Marie Sklodowska-Curie grant agreement No. 847523 ``INTERACTIONS''.
  This work is supported by World Premier International Research Center Initiative (WPI Initiative), MEXT, Japan, as well as KAKENHI Grant-in-Aid for Scientific Research (A) (15H02064, 17H01110, 17H01114, 20H00180, and 21H04467) through Japan Society for the Promotion of Science (JSPS). 
  This work has been supported in part by JSPS KAKENHI Grant Nos. JP17K05382, JP20K04024, and JP21H04499 (K.N.).
  Yuki Isobe, Kimihiko Nakajima, Yuichi Harikane, Takashi Kojima, and Masato Onodera are supported by JSPS KAKENHI Grant Nos. 21J20785, 20K22373, 19J01222, 18J12840, and 17K14257, respectively.


%% \bibliography{Isobe+20}{}
\bibliography{library}

\begin{thebibliography}{}
\expandafter\ifx\csname natexlab\endcsname\relax\def\natexlab#1{#1}\fi
\providecommand{\url}[1]{\href{#1}{#1}}
\providecommand{\dodoi}[1]{doi:~\href{http://doi.org/#1}{\nolinkurl{#1}}}
\providecommand{\doeprint}[1]{\href{http://ascl.net/#1}{\nolinkurl{http://ascl.net/#1}}}
\providecommand{\doarXiv}[1]{\href{https://arxiv.org/abs/#1}{\nolinkurl{https://arxiv.org/abs/#1}}}

\bibitem[{Aihara {et~al.}(2019)Aihara, AlSayyad, Ando, Armstrong, Bosch, Egami,
  Furusawa, Furusawa, Goulding, Harikane, Hikage, Ho, Hsieh, Huang, Ikeda,
  Imanishi, Ito, Iwata, Jaelani, Kakuma, Kawana, Kikuta, Kobayashi, Koike,
  Komiyama, Li, Liang, Lin, Luo, Lupton, Lust, MacArthur, Matsuoka, Mineo,
  Miyatake, Miyazaki, More, Murata, Namiki, Nishizawa, Oguri, Okabe, Okamoto,
  Okura, Ono, Onodera, Onoue, Osato, Ouchi, Shibuya, Strauss, Sugiyama, Suto,
  Takada, Takagi, Takata, Takita, Tanaka, Terai, Toba, Uchiyama, Utsumi, Wang,
  Wang, \& Yamada}]{Aihara2019}
Aihara, H., AlSayyad, Y., Ando, M., {et~al.} 2019, PASJ,
  \dodoi{10.1093/pasj/psz103}

\bibitem[{Andrews \& Martini(2013)}]{Andrews2013}
Andrews, B.~H., \& Martini, P. 2013, ApJ, \dodoi{10.1088/0004-637X/765/2/140}

\bibitem[{{Aoki} {et~al.}(2014){Aoki}, {Tominaga}, {Beers}, {Honda}, \&
  {Lee}}]{Aoki2014}
{Aoki}, W., {Tominaga}, N., {Beers}, T.~C., {Honda}, S., \& {Lee}, Y.~S. 2014,
  Science, 345, 912, \dodoi{10.1126/science.1252633}

\bibitem[{{Asplund} {et~al.}(2021){Asplund}, {Amarsi}, \&
  {Grevesse}}]{Asplund2021}
{Asplund}, M., {Amarsi}, A.~M., \& {Grevesse}, N. 2021, \aap, 653, A141,
  \dodoi{10.1051/0004-6361/202140445}

\bibitem[{{Becker} {et~al.}(2012){Becker}, {Sargent}, {Rauch}, \&
  {Carswell}}]{Becker2012}
{Becker}, G.~D., {Sargent}, W. L.~W., {Rauch}, M., \& {Carswell}, R.~F. 2012,
  \apj, 744, 91, \dodoi{10.1088/0004-637X/744/2/91}

\bibitem[{Behroozi {et~al.}(2013)Behroozi, Wechsler, \& Conroy}]{Behroozi2013}
Behroozi, P.~S., Wechsler, R.~H., \& Conroy, C. 2013, ApJ, 770,
  \dodoi{10.1088/0004-637X/770/1/57}

\bibitem[{{B{\`e}land} {et~al.}(1988){B{\`e}land}, {Boulade}, \&
  {Davidge}}]{Beland1988}
{B{\`e}land}, S., {Boulade}, O., \& {Davidge}, T. 1988, Bulletin d'information
  du telescope Canada-France-Hawaii, 19, 16

\bibitem[{{Bensby} {et~al.}(2014){Bensby}, {Feltzing}, \& {Oey}}]{Bensby2014}
{Bensby}, T., {Feltzing}, S., \& {Oey}, M.~S. 2014, \aap, 562, A71,
  \dodoi{10.1051/0004-6361/201322631}

\bibitem[{Berg {et~al.}(2019)Berg, Erb, Henry, Skillman, \& McQuinn}]{Berg2019}
Berg, D.~A., Erb, D.~K., Henry, R. B.~C., Skillman, E.~D., \& McQuinn, K. B.~W.
  2019, ApJ, 874, 93, \dodoi{10.3847/1538-4357/ab020a}

\bibitem[{{Berg} {et~al.}(2015){Berg}, {Skillman}, {Croxall}, {Pogge},
  {Moustakas}, \& {Johnson-Groh}}]{Berg2015}
{Berg}, D.~A., {Skillman}, E.~D., {Croxall}, K.~V., {et~al.} 2015, \apj, 806,
  16, \dodoi{10.1088/0004-637X/806/1/16}

\bibitem[{Bruzual \& Charlot(2003)}]{Bruzual2003}
Bruzual, G., \& Charlot, S. 2003, MNRAS,
  \dodoi{10.1046/j.1365-8711.2003.06897.x}

\bibitem[{Calzetti {et~al.}(2000)Calzetti, Armus, Bohlin, Kinney, Koornneef, \&
  Storchi‐Bergmann}]{Calzetti2000}
Calzetti, D., Armus, L., Bohlin, R.~C., {et~al.} 2000, ApJ, 533, 682,
  \dodoi{10.1086/308692}

\bibitem[{{Calzetti} {et~al.}(1994){Calzetti}, {Kinney}, \&
  {Storchi-Bergmann}}]{Calzetti1994}
{Calzetti}, D., {Kinney}, A.~L., \& {Storchi-Bergmann}, T. 1994, \apj, 429,
  582, \dodoi{10.1086/174346}

\bibitem[{{Cayrel} {et~al.}(2004){Cayrel}, {Depagne}, {Spite}, {Hill}, {Spite},
  {Fran{\c{c}}ois}, {Plez}, {Beers}, {Primas}, {Andersen}, {Barbuy},
  {Bonifacio}, {Molaro}, \& {Nordstr{\"o}m}}]{Cayrel2004}
{Cayrel}, R., {Depagne}, E., {Spite}, M., {et~al.} 2004, \aap, 416, 1117,
  \dodoi{10.1051/0004-6361:20034074}

\bibitem[{Chabrier(2003)}]{Chabrier2003}
Chabrier, G. 2003, PASP, \dodoi{10.1086/376392}

\bibitem[{Chevallard \& Charlot(2016)}]{Chevallard2016}
Chevallard, J., \& Charlot, S. 2016, MNRAS, 462, 1415,
  \dodoi{10.1093/mnras/stw1756}

\bibitem[{{Edvardsson} {et~al.}(1993){Edvardsson}, {Andersen}, {Gustafsson},
  {Lambert}, {Nissen}, \& {Tomkin}}]{Edvardsson1993}
{Edvardsson}, B., {Andersen}, J., {Gustafsson}, B., {et~al.} 1993, \aap, 500,
  391

\bibitem[{Ferland {et~al.}(2013)Ferland, Porter, {Van Hoof}, Williams, Abel,
  Lykins, Shaw, Henney, \& Stancil}]{Ferland2013}
Ferland, G.~J., Porter, R.~L., {Van Hoof}, P.~A., {et~al.} 2013, {The 2013
  release of CLOUDY}.
\newblock \doarXiv{1302.4485}

\bibitem[{{Froese Fischer} \& {Tachiev}(2004)}]{FroeseFischer2004}
{Froese Fischer}, C., \& {Tachiev}, G. 2004, Atomic Data and Nuclear Data
  Tables, 87, 1, \dodoi{10.1016/j.adt.2004.02.001}

\bibitem[{Gardner {et~al.}(2006)Gardner, Mather, Clampin, Doyon, Greenhouse,
  Hammel, Hutchings, Jakobsen, Lilly, Long, Lunine, McCaughrean, Mountain,
  Nella, Rieke, Rieke, Rix, Smith, Sonneborn, Stiavelli, Stockman, Windhorst,
  \& Wright}]{Gardner2006}
Gardner, J.~P., Mather, J.~C., Clampin, M., {et~al.} 2006, Space Science
  Reviews, 123, 485, \dodoi{10.1007/s11214-006-8315-7}

\bibitem[{Garnett(1992)}]{Garnett1992}
Garnett, D.~R. 1992, AJ, \dodoi{10.1086/116146}

\bibitem[{{Gratton} {et~al.}(2003){Gratton}, {Carretta}, {Claudi}, {Lucatello},
  \& {Barbieri}}]{Gratton2003}
{Gratton}, R.~G., {Carretta}, E., {Claudi}, R., {Lucatello}, S., \& {Barbieri},
  M. 2003, \aap, 404, 187, \dodoi{10.1051/0004-6361:20030439}

\bibitem[{Gutkin {et~al.}(2016)Gutkin, Charlot, \& Bruzual}]{Gutkin2016}
Gutkin, J., Charlot, S., \& Bruzual, G. 2016, MNRAS,
  \dodoi{10.1093/mnras/stw1716}

\bibitem[{Hees {et~al.}(2015)Hees, Hestroffer, Poncin-Lafitte, \&
  David}]{Hees2015}
Hees, a., Hestroffer, D., Poncin-Lafitte, C.~L., \& David, P. 2015, Research in
  Astronomy and Astrophysics, 15, 1945.
\newblock \doarXiv{1509.06868}

\bibitem[{Heger \& Woosley(2002)}]{Heger2002}
Heger, A., \& Woosley, S.~E. 2002, ApJ, \dodoi{10.1086/338487}

\bibitem[{Hirano {et~al.}(2015)Hirano, Hosokawa, Yoshida, Omukai, \&
  Yorke}]{Hirano2015}
Hirano, S., Hosokawa, T., Yoshida, N., Omukai, K., \& Yorke, H.~W. 2015, MNRAS,
  448, 568, \dodoi{10.1093/mnras/stv044}

\bibitem[{Hirano {et~al.}(2014)Hirano, Hosokawa, Yoshida, Umeda, Omukai,
  Chiaki, \& Yorke}]{Hirano2014}
Hirano, S., Hosokawa, T., Yoshida, N., {et~al.} 2014, ApJ,
  \dodoi{10.1088/0004-637X/781/2/60}

\bibitem[{Hirschauer {et~al.}(2016)Hirschauer, Salzer, Skillman, Berg, McQuinn,
  Cannon, Gordon, Haynes, Giovanelli, Adams, Janowiecki, Rhode, Pogge, Croxall,
  \& Aver}]{Hirschauer2016}
Hirschauer, A.~S., Salzer, J.~J., Skillman, E.~D., {et~al.} 2016, ApJ,
  \dodoi{10.3847/0004-637x/822/2/108}

\bibitem[{Hsyu {et~al.}(2017)Hsyu, Cooke, Prochaska, \& Bolte}]{Hsyu2017}
Hsyu, T., Cooke, R.~J., Prochaska, J.~X., \& Bolte, M. 2017, ApJ,
  \dodoi{10.3847/2041-8213/aa821f}

\bibitem[{Hsyu {et~al.}(2018)Hsyu, Cooke, Prochaska, \& Bolte}]{Hsyu2018}
---. 2018, ApJ, 863, 134, \dodoi{10.3847/1538-4357/aad18a}

\bibitem[{{Huang} {et~al.}(2018){Huang}, {Leauthaud}, {Murata}, {Bosch},
  {Price}, {Lupton}, {Mandelbaum}, {Lackner}, {Bickerton}, {Miyazaki},
  {Coupon}, \& {Tanaka}}]{Huang2018}
{Huang}, S., {Leauthaud}, A., {Murata}, R., {et~al.} 2018, \pasj, 70, S6,
  \dodoi{10.1093/pasj/psx126}

\bibitem[{{Hughes} {et~al.}(2013){Hughes}, {Cortese}, {Boselli}, {Gavazzi}, \&
  {Davies}}]{Hughes2013}
{Hughes}, T.~M., {Cortese}, L., {Boselli}, A., {Gavazzi}, G., \& {Davies},
  J.~I. 2013, \aap, 550, A115, \dodoi{10.1051/0004-6361/201218822}

\bibitem[{Inoue(2011)}]{Inoue2011}
Inoue, A.~K. 2011, MNRAS, \dodoi{10.1111/j.1365-2966.2011.18906.x}

\bibitem[{Isobe {et~al.}(2021)Isobe, Ouchi, Kojima, Shibuya, Hayashi, Rauch,
  Kikuchihara, Zhang, Ono, Fujimoto, Harikane, Kim, Komiyama, Kusakabe, Lee,
  Mawatari, Onodera, Sugahara, \& Yabe}]{Isobe2020}
Isobe, Y., Ouchi, M., Kojima, T., {et~al.} 2021, \apj \ in Press.
\newblock \doarXiv{2004.11444}

\bibitem[{{Iwamoto} {et~al.}(1998){Iwamoto}, {Mazzali}, {Nomoto}, {Umeda},
  {Nakamura}, {Patat}, {Danziger}, {Young}, {Suzuki}, {Shigeyama},
  {Augusteijn}, {Doublier}, {Gonzalez}, {Boehnhardt}, {Brewer}, {Hainaut},
  {Lidman}, {Leibundgut}, {Cappellaro}, {Turatto}, {Galama}, {Vreeswijk},
  {Kouveliotou}, {van Paradijs}, {Pian}, {Palazzi}, \&
  {Frontera}}]{Iwamoto1998}
{Iwamoto}, K., {Mazzali}, P.~A., {Nomoto}, K., {et~al.} 1998, \nat, 395, 672,
  \dodoi{10.1038/27155}

\bibitem[{Izotov {et~al.}(2009)Izotov, Guseva, Fricke, \&
  Papaderos}]{Izotov2009}
Izotov, Y.~I., Guseva, N.~G., Fricke, K.~J., \& Papaderos, P. 2009, A{\&}A,
  \dodoi{10.1051/0004-6361/200911965}

\bibitem[{Izotov {et~al.}(2006)Izotov, Stasi{\'{n}}ska, Meynet, Guseva, \&
  Thuan}]{Izotov2006}
Izotov, Y.~I., Stasi{\'{n}}ska, G., Meynet, G., Guseva, N.~G., \& Thuan, T.~X.
  2006, A{\&}A, 448, 955, \dodoi{10.1051/0004-6361:20053763}

\bibitem[{Izotov \& Thuan(1998)}]{Izotov1998}
Izotov, Y.~I., \& Thuan, T.~X. 1998, ApJ, 497, 227, \dodoi{10.1086/305440}

\bibitem[{Izotov {et~al.}(2019)Izotov, Thuan, \& Guseva}]{Izotov2019}
Izotov, Y.~I., Thuan, T.~X., \& Guseva, N.~G. 2019, MNRAS, 483, 5491,
  \dodoi{10.1093/mnras/sty3472}

\bibitem[{{Izotov} {et~al.}(2021){Izotov}, {Thuan}, \& {Guseva}}]{Izotov2021a}
{Izotov}, Y.~I., {Thuan}, T.~X., \& {Guseva}, N.~G. 2021, \mnras, 504, 3996,
  \dodoi{10.1093/mnras/stab1099}

\bibitem[{Izotov {et~al.}(2018)Izotov, Worseck, Schaerer, Guseva, Thuan,
  Fricke, Verhamme, \& Orlitov{\'{a}}}]{Izotov2018}
Izotov, Y.~I., Worseck, G., Schaerer, D., {et~al.} 2018, MNRAS, 478, 4851,
  \dodoi{10.1093/mnras/sty1378}

\bibitem[{{James} {et~al.}(2015){James}, {Koposov}, {Stark}, {Belokurov},
  {Pettini}, \& {Olszewski}}]{James2015}
{James}, B.~L., {Koposov}, S., {Stark}, D.~P., {et~al.} 2015, \mnras, 448,
  2687, \dodoi{10.1093/mnras/stv175}

\bibitem[{James {et~al.}(2017)James, Koposov, Stark, Belokurov, Pettini,
  Olszewski, \& McQuinn}]{James2017}
James, B.~L., Koposov, S.~E., Stark, D.~P., {et~al.} 2017, MNRAS, 465, 3977,
  \dodoi{10.1093/mnras/stw2962}

\bibitem[{{Johansson} {et~al.}(2000){Johansson}, {Zethson}, {Hartman},
  {Ekberg}, {Ishibashi}, {Davidson}, \& {Gull}}]{Johansson2000}
{Johansson}, S., {Zethson}, T., {Hartman}, H., {et~al.} 2000, \aap, 361, 977

\bibitem[{Kikuchihara {et~al.}(2020)Kikuchihara, Ouchi, Ono, Mawatari,
  Chevallard, Harikane, Kojima, Oguri, Bruzual, \& Charlot}]{Kikuchihara2020}
Kikuchihara, S., Ouchi, M., Ono, Y., {et~al.} 2020, ApJ,
  \dodoi{10.3847/1538-4357/ab7dbe}

\bibitem[{{Kisielius} {et~al.}(2009){Kisielius}, {Storey}, {Ferland}, \&
  {Keenan}}]{Kisielius2009}
{Kisielius}, R., {Storey}, P.~J., {Ferland}, G.~J., \& {Keenan}, F.~P. 2009,
  \mnras, 397, 903, \dodoi{10.1111/j.1365-2966.2009.14989.x}

\bibitem[{Kojima {et~al.}(2020)Kojima, Ouchi, Rauch, Ono, Nakajima, Isobe,
  Fujimoto, Harikane, Hashimoto, Hayashi, Komiyama, Kusakabe, Kim, Lee, Mukae,
  Nagao, Onodera, Shibuya, Sugahara, Umemura, \& Yabe}]{Kojima2020a}
Kojima, T., Ouchi, M., Rauch, M., {et~al.} 2020, ApJ, 898, 142,
  \dodoi{10.3847/1538-4357/aba047}

\bibitem[{{Kojima} {et~al.}(2021){Kojima}, {Ouchi}, {Rauch}, {Ono}, {Nakajima},
  {Isobe}, {Fujimoto}, {Harikane}, {Hashimoto}, {Hayashi}, {Komiyama},
  {Kusakabe}, {Kim}, {Lee}, {Mukae}, {Nagao}, {Onodera}, {Shibuya}, {Sugahara},
  {Umemura}, \& {Yabe}}]{Kojima2021}
{Kojima}, T., {Ouchi}, M., {Rauch}, M., {et~al.} 2021, \apj, 913, 22,
  \dodoi{10.3847/1538-4357/abec3d}

\bibitem[{Kroupa(2001)}]{Kroupa2001}
Kroupa, P. 2001, MNRAS, \dodoi{10.1046/j.1365-8711.2001.04022.x}

\bibitem[{{Kumari} {et~al.}(2018){Kumari}, {James}, {Irwin}, {Amor{\'\i}n}, \&
  {P{\'e}rez-Montero}}]{Kumari2018}
{Kumari}, N., {James}, B.~L., {Irwin}, M.~J., {Amor{\'\i}n}, R., \&
  {P{\'e}rez-Montero}, E. 2018, \mnras, 476, 3793, \dodoi{10.1093/mnras/sty402}

\bibitem[{{Langer} {et~al.}(2007){Langer}, {Norman}, {de Koter}, {Vink},
  {Cantiello}, \& {Yoon}}]{Langer2007}
{Langer}, N., {Norman}, C.~A., {de Koter}, A., {et~al.} 2007, \aap, 475, L19,
  \dodoi{10.1051/0004-6361:20078482}

\bibitem[{{Lilly} {et~al.}(2013){Lilly}, {Carollo}, {Pipino}, {Renzini}, \&
  {Peng}}]{Lilly2013}
{Lilly}, S.~J., {Carollo}, C.~M., {Pipino}, A., {Renzini}, A., \& {Peng}, Y.
  2013, \apj, 772, 119, \dodoi{10.1088/0004-637X/772/2/119}

\bibitem[{{Limongi} \& {Chieffi}(2006)}]{Limongi2006}
{Limongi}, M., \& {Chieffi}, A. 2006, \apj, 647, 483, \dodoi{10.1086/505164}

\bibitem[{{Luridiana} {et~al.}(2015){Luridiana}, {Morisset}, \&
  {Shaw}}]{Luridiana2015}
{Luridiana}, V., {Morisset}, C., \& {Shaw}, R.~A. 2015, \aap, 573, A42,
  \dodoi{10.1051/0004-6361/201323152}

\bibitem[{Mannucci {et~al.}(2005)Mannucci, Valle, \& Panagia}]{Mannucci2005}
Mannucci, F., Valle, M.~D., \& Panagia, N. 2005, arXiv

\bibitem[{{McLaughlin} \& {Bell}(2000)}]{McLaughlin2000}
{McLaughlin}, B.~M., \& {Bell}, K.~L. 2000, JPhB, 33, 597,
  \dodoi{10.1088/0953-4075/33/4/301}

\bibitem[{{Modjaz} {et~al.}(2020){Modjaz}, {Bianco}, {Siwek}, {Huang},
  {Perley}, {Fierroz}, {Liu}, {Arcavi}, {Gal-Yam}, {Filippenko},
  {Blagorodnova}, {Cenko}, {Kasliwal}, {Kulkarni}, {Schulze}, {Taggart}, \&
  {Zheng}}]{Maryam2020}
{Modjaz}, M., {Bianco}, F.~B., {Siwek}, M., {et~al.} 2020, \apj, 892, 153,
  \dodoi{10.3847/1538-4357/ab4185}

\bibitem[{Morales-Luis {et~al.}(2011)Morales-Luis, {S{\'{a}}nchez Almeida},
  Aguerri, \& Mũoz-Tũ{\'{o}}n}]{Morales-Luis2011}
Morales-Luis, A.~B., {S{\'{a}}nchez Almeida}, J., Aguerri, J.~A., \&
  Mũoz-Tũ{\'{o}}n, C. 2011, ApJ, 743, \dodoi{10.1088/0004-637X/743/1/77}

\bibitem[{{Munoz Burgos} {et~al.}(2009){Munoz Burgos}, {Loch}, {Ballance}, \&
  {Boivin}}]{MunozBurgos2009}
{Munoz Burgos}, J.~M., {Loch}, S.~D., {Ballance}, C.~P., \& {Boivin}, R.~F.
  2009, \aap, 500, 1253, \dodoi{10.1051/0004-6361/200911743}

\bibitem[{{Nadyozhin}(1994)}]{Nadyozhin1994}
{Nadyozhin}, D.~K. 1994, ApJS, 92, 527, \dodoi{10.1086/192008}

\bibitem[{{Nakamura} {et~al.}(2001){Nakamura}, {Mazzali}, {Nomoto}, \&
  {Iwamoto}}]{Nakamura2001}
{Nakamura}, T., {Mazzali}, P.~A., {Nomoto}, K., \& {Iwamoto}, K. 2001, \apj,
  550, 991, \dodoi{10.1086/319784}

\bibitem[{{Nomoto} {et~al.}(2013){Nomoto}, {Kobayashi}, \&
  {Tominaga}}]{Nomoto2013}
{Nomoto}, K., {Kobayashi}, C., \& {Tominaga}, N. 2013, \araa, 51, 457,
  \dodoi{10.1146/annurev-astro-082812-140956}

\bibitem[{{Nomoto} {et~al.}(1984){Nomoto}, {Thielemann}, \&
  {Yokoi}}]{Nomoto1984}
{Nomoto}, K., {Thielemann}, F.~K., \& {Yokoi}, K. 1984, \apj, 286, 644,
  \dodoi{10.1086/162639}

\bibitem[{{Nomoto} {et~al.}(2006){Nomoto}, {Tominaga}, {Umeda}, {Kobayashi}, \&
  {Maeda}}]{Nomoto2006}
{Nomoto}, K., {Tominaga}, N., {Umeda}, H., {Kobayashi}, C., \& {Maeda}, K.
  2006, \nphysa, 777, 424, \dodoi{10.1016/j.nuclphysa.2006.05.008}

\bibitem[{Oke \& Gunn(1983)}]{Oke1983}
Oke, J.~B., \& Gunn, J.~E. 1983, ApJ, \dodoi{10.1086/160817}

\bibitem[{Oke {et~al.}(1995)Oke, Cohen, Carr, Cromer, Dingizian, Harris,
  Labrecque, Lucinio, Schaal, Epps, \& Miller}]{Oke1995}
Oke, J.~B., Cohen, J.~G., Carr, M., {et~al.} 1995, PASP, 107, 375,
  \dodoi{10.1086/133562}

\bibitem[{Ono {et~al.}(2018)Ono, Ouchi, Harikane, Toshikawa, Rauch, Yuma,
  Sawicki, Shibuya, Shimasaku, Oguri, Willott, Akhlaghi, Akiyama, Coupon,
  Kashikawa, Komiyama, Konno, Lin, Matsuoka, Miyazaki, Nagao, Nakajima,
  Silverman, Tanaka, Taniguchi, \& Wang}]{Ono2018}
Ono, Y., Ouchi, M., Harikane, Y., {et~al.} 2018, PASJ,
  \dodoi{10.1093/pasj/psx103}

\bibitem[{Padovani \& Matteucci(1993)}]{Padovani1993}
Padovani, P., \& Matteucci, F. 1993, ApJ, \dodoi{10.1086/173212}

\bibitem[{{Pomp{\'e}ia} {et~al.}(2008){Pomp{\'e}ia}, {Hill}, {Spite}, {Cole},
  {Primas}, {Romaniello}, {Pasquini}, {Cioni}, \& {Smecker Hane}}]{Pompeia2008}
{Pomp{\'e}ia}, L., {Hill}, V., {Spite}, M., {et~al.} 2008, \aap, 480, 379,
  \dodoi{10.1051/0004-6361:20064854}

\bibitem[{{Portinari} {et~al.}(1998){Portinari}, {Chiosi}, \&
  {Bressan}}]{Portinari1998}
{Portinari}, L., {Chiosi}, C., \& {Bressan}, A. 1998, \aap, 334, 505.
\newblock \doarXiv{astro-ph/9711337}

\bibitem[{{Prochaska} {et~al.}(2003){Prochaska}, {Gawiser}, {Wolfe}, {Castro},
  \& {Djorgovski}}]{Prochaska2003}
{Prochaska}, J.~X., {Gawiser}, E., {Wolfe}, A.~M., {Castro}, S., \&
  {Djorgovski}, S.~G. 2003, \apj, 595, L9.
\newblock \url{https://iopscience.iop.org/article/10.1086/378945}

\bibitem[{Pustilnik {et~al.}(2005)Pustilnik, Kniazev, \&
  Pramskij}]{Pustilnik2005}
Pustilnik, S.~A., Kniazev, A.~Y., \& Pramskij, A.~G. 2005, A{\&}A,
  \dodoi{10.1051/0004-6361:20053102}

\bibitem[{{Quinet}(1996)}]{Quinet1996}
{Quinet}, P. 1996, \aaps, 116, 573

\bibitem[{{Reddy} {et~al.}(2003){Reddy}, {Tomkin}, {Lambert}, \& {Allende
  Prieto}}]{Reddy2003}
{Reddy}, B.~E., {Tomkin}, J., {Lambert}, D.~L., \& {Allende Prieto}, C. 2003,
  \mnras, 340, 304, \dodoi{10.1046/j.1365-8711.2003.06305.x}

\bibitem[{Rodriguez \& Rubin(2005)}]{Rodriguez2005}
Rodriguez, M., \& Rubin, R.~H. 2005, ApJ, \dodoi{10.1086/429958}

\bibitem[{{Roederer} {et~al.}(2014){Roederer}, {Preston}, {Thompson},
  {Shectman}, {Sneden}, {Burley}, \& {Kelson}}]{Roederer2014}
{Roederer}, I.~U., {Preston}, G.~W., {Thompson}, I.~B., {et~al.} 2014, \aj,
  147, 136, \dodoi{10.1088/0004-6256/147/6/136}

\bibitem[{{Romano} {et~al.}(2010){Romano}, {Karakas}, {Tosi}, \&
  {Matteucci}}]{Romano2010}
{Romano}, D., {Karakas}, A.~I., {Tosi}, M., \& {Matteucci}, F. 2010, \aap, 522,
  A32, \dodoi{10.1051/0004-6361/201014483}

\bibitem[{Sacchi {et~al.}(2016)Sacchi, Annibali, Cignoni, Aloisi, Sohn, Tosi,
  van~der Marel, Grocholski, \& James}]{Sacchi2016}
Sacchi, E., Annibali, F., Cignoni, M., {et~al.} 2016, ApJ, 830, 3,
  \dodoi{10.3847/0004-637x/830/1/3}

\bibitem[{Salpeter(1955)}]{Salpeter1955}
Salpeter, E.~E. 1955, ApJ, \dodoi{10.1086/145971}

\bibitem[{{S{\'{a}}nchez Almeida} {et~al.}(2016){S{\'{a}}nchez Almeida},
  Perez-Montero, Morales-Luis, Munoz-Tunon, Garcia-Benito, Nuza, \&
  Kitaura}]{Almeida2016}
{S{\'{a}}nchez Almeida}, J., Perez-Montero, E., Morales-Luis, A.~B., {et~al.}
  2016, \apj, 819, 110, \dodoi{10.3847/0004-637X/819/2/110}

\bibitem[{{S{\'{a}}nchez Almeida} {et~al.}(2015){S{\'{a}}nchez Almeida},
  Elmegreen, Mu{\~{n}}oz-Tu{\'{o}}n, Elmegreen, P{\'{e}}rez-Montero,
  Amor{\'{i}}n, Filho, Ascasibar, Papaderos, \&
  V{\'{i}}lchez}]{SanchezAlmeida2015}
{S{\'{a}}nchez Almeida}, J., Elmegreen, B.~G., Mu{\~{n}}oz-Tu{\'{o}}n, C.,
  {et~al.} 2015, ApJL, 810, L15, \dodoi{10.1088/2041-8205/810/2/L15}

\bibitem[{Schlafly \& Finkbeiner(2011)}]{Schlafly2011}
Schlafly, E.~F., \& Finkbeiner, D.~P. 2011, ApJ, 737,
  \dodoi{10.1088/0004-637X/737/2/103}

\bibitem[{Senchyna \& Stark(2019)}]{Senchyna2019}
Senchyna, P., \& Stark, D.~P. 2019, MNRAS, 484, 1270,
  \dodoi{10.1093/mnras/stz058}

\bibitem[{{Shibata} \& {Shapiro}(2002)}]{Shibata2002}
{Shibata}, M., \& {Shapiro}, S.~L. 2002, \apjl, 572, L39,
  \dodoi{10.1086/341516}

\bibitem[{Shivvers {et~al.}(2017)Shivvers, Modjaz, Zheng, Liu, Filippenko,
  Silverman, Matheson, Pastorello, Graur, Foley, Chornock, Smith, Leaman, \&
  Benetti}]{Shivvers2017}
Shivvers, I., Modjaz, M., Zheng, W., {et~al.} 2017, PASP, 129, 54201,
  \dodoi{10.1088/1538-3873/aa54a6}

\bibitem[{Skillman {et~al.}(2013)Skillman, Salzer, Berg, Pogge, Haurberg,
  Cannon, Aver, Olive, Giovanelli, Haynes, Adams, McQuinn, \&
  Rhode}]{Skillman2013}
Skillman, E.~D., Salzer, J.~J., Berg, D.~A., {et~al.} 2013, AJ,
  \dodoi{10.1088/0004-6256/146/1/3}

\bibitem[{{Storey} \& {Hummer}(1995)}]{Storey1995}
{Storey}, P.~J., \& {Hummer}, D.~G. 1995, \mnras, 272, 41,
  \dodoi{10.1093/mnras/272.1.41}

\bibitem[{{Storey} {et~al.}(2014){Storey}, {Sochi}, \& {Badnell}}]{Storey2014}
{Storey}, P.~J., {Sochi}, T., \& {Badnell}, N.~R. 2014, \mnras, 441, 3028,
  \dodoi{10.1093/mnras/stu777}

\bibitem[{Sullivan {et~al.}(2006)Sullivan, {Le Borgne}, Pritchet, Hodsman,
  Neill, Howell, Carlberg, Astier, Aubourg, Balam, Basa, Conley, Fabbro,
  Fouchez, Guy, Hook, Pain, Palanque‐Delabrouille, Perrett, Regnault, Rich,
  Taillet, Baumont, Bronder, Ellis, Filiol, Lusset, Perlmutter, Ripoche, \&
  Tao}]{Sullivan2006}
Sullivan, M., {Le Borgne}, D., Pritchet, C.~J., {et~al.} 2006, ApJ,
  \dodoi{10.1086/506137}

\bibitem[{Suzuki \& Maeda(2018)}]{Suzuki2018a}
Suzuki, A., \& Maeda, K. 2018, \apj, 852, 101, \dodoi{10.3847/1538-4357/aaa024}

\bibitem[{Takahashi {et~al.}(2018)Takahashi, Yoshida, \& Umeda}]{Takahashi2018}
Takahashi, K., Yoshida, T., \& Umeda, H. 2018, arXiv,
  \dodoi{10.3847/1538-4357/aab95f}

\bibitem[{{Tayal}(2011)}]{Tayal2011}
{Tayal}, S.~S. 2011, \apjs, 195, 12, \dodoi{10.1088/0067-0049/195/2/12}

\bibitem[{{Thuan} {et~al.}(2005){Thuan}, {Lecavelier des Etangs}, \&
  {Izotov}}]{Thuan2005}
{Thuan}, T.~X., {Lecavelier des Etangs}, A., \& {Izotov}, Y.~I. 2005, \apj,
  621, 269, \dodoi{10.1086/427469}

\bibitem[{Totani {et~al.}(2008)Totani, Morokuma, Oda, Doi, \&
  Yasuda}]{Totani2008}
Totani, T., Morokuma, T., Oda, T., Doi, M., \& Yasuda, N. 2008, PASJ, 60, 1327,
  \dodoi{10.1093/pasj/60.6.1327}

\bibitem[{Tumlinson \& Shull(2000)}]{Tumlinson2000}
Tumlinson, J., \& Shull, J.~M. 2000, ApJ, 528, L65, \dodoi{10.1086/312432}

\bibitem[{{Umeda} \& {Nomoto}(2008)}]{Umeda2008}
{Umeda}, H., \& {Nomoto}, K. 2008, \apj, 673, 1014, \dodoi{10.1086/524767}

\bibitem[{Vincenzo {et~al.}(2016)Vincenzo, Belfiore, Maiolino, Matteucci, \&
  Ventura}]{Vincenzo2016}
Vincenzo, F., Belfiore, F., Maiolino, R., Matteucci, F., \& Ventura, P. 2016,
  MNRAS, 458, 3466, \dodoi{10.1093/mnras/stw532}

\bibitem[{{Vink}(2018)}]{Vink2018}
{Vink}, J.~S. 2018, \aap, 615, A119, \dodoi{10.1051/0004-6361/201832773}

\bibitem[{Wise {et~al.}(2012)Wise, Turk, Norman, \& Abel}]{Wise2012a}
Wise, J.~H., Turk, M.~J., Norman, M.~L., \& Abel, T. 2012, ApJ, 745,
  \dodoi{10.1088/0004-637X/745/1/50}

\bibitem[{{Xing} {et~al.}(2019){Xing}, {Zhao}, {Aoki}, {Honda}, {Li},
  {Ishigaki}, \& {Matsuno}}]{Xing2019}
{Xing}, Q.-F., {Zhao}, G., {Aoki}, W., {et~al.} 2019, Nature Astronomy, 3, 631,
  \dodoi{10.1038/s41550-019-0764-5}

\bibitem[{{Zahid} {et~al.}(2012){Zahid}, {Bresolin}, {Kewley}, {Coil}, \&
  {Dav{\'e}}}]{Zahid2012}
{Zahid}, H.~J., {Bresolin}, F., {Kewley}, L.~J., {Coil}, A.~L., \& {Dav{\'e}},
  R. 2012, \apj, 750, 120, \dodoi{10.1088/0004-637X/750/2/120}

\bibitem[{{Zhang}(1996)}]{Zhang1996}
{Zhang}, H. 1996, AAS, 119, 523, \dodoi{10.1051/aas:1996264}

\end{thebibliography}
%\bibliographystyle{apj}


\end{document}